\documentclass[twocolumn]{aastex631}

\shorttitle{probing early reionization}
\shortauthors{Sakamoto et al.}
\graphicspath{{./}{figures/}}
\usepackage{comment}
\usepackage{amsmath}
\begin{document}


\title[Probing early history of reionization]{Probing the Early History of Cosmic Reionization by Future Cosmic Microwave Background Experiments}

\correspondingauthor{Kyungjin Ahn}
\email{sakamoto.hina@nagoya-u.jp, kjahn@chosun.ac.kr}

\author{Hina Sakamoto}
\affiliation{Department of Physics, Graduate School of Science, Nagoya University, Aichi 464-8602, Japan}

\author{Kyungjin Ahn}
\affiliation{Department of Earth Sciences, Chosun University, Gwangju 61452, Republic of Korea}

\author{Kiyotomo Ichiki}
\affiliation{Department of Physics, Graduate School of Science, Nagoya University, Aichi 464-8602, Japan}
\affiliation{Kobayashi-Maskawa Institute for the Origin of Particles and the Universe, Nagoya University, Aichi 464-8602, Japan}

\author{Hyunjin Moon}
\affiliation{Department of Earth Sciences, Chosun University, Gwangju 61452, Republic of Korea}

\author{Kenji Hasegawa}
\affiliation{Department of Physics, Graduate School of Science, Nagoya University, Aichi 464-8602, Japan}

\def\mbf#1{\mbox{\boldmath ${#1}$}} 

\newcommand{\del}{\partial}
\newcommand{\xe}{x_{\mathrm{e}}}
\newcommand{\ClTT}{C_\ell^{TT}}
\newcommand{\ClTE}{C_\ell^{TE}}
\newcommand{\ClEE}{C_\ell^{EE}}
\newcommand{\xefid}{x_\mathrm{e,fid}}
\newcommand{\tft}{\tau_{z>15}}
\newcommand{\KI}[1]{\textcolor{red}{\bf #1}}
\newcommand{\KA}[1]{\textcolor{blue}{\bf #1}}
\newcommand{\HS}[1]{\textcolor{magenta}{\bf #1}}



\begin{abstract}
Cosmic Reionization imprints its signature on the temperature and polarization anisotropies of the cosmic microwave background (CMB). Advances in CMB telescopes have already placed a significant constraint on the history of reionization. As near-future CMB telescopes target the maximum sensitivity, or observations limited only by the cosmic variance (CV), we hereby forecast the potential of future CMB observations in constraining the history of reionization. In this study, we perform Markov Chain Monte Carlo analysis for CV-limited E-mode polarization observations such as the LiteBIRD (Light satellite for the studies of B-mode polarization and Inflation from cosmic background Radiation Detection), based on a few different methods that vary in the way of sampling reionization histories. We focus especially on estimating the very early history of reionization that occurs at redshifts $z>15$, which is quantified by the partial CMB optical depth due to free electrons at $z>15$, $\tft$. We find that reionization with $\tft \sim 0.008$, which are well below the current upper limit $\tft \sim 0.02$, are achievable by reionization models with minihalo domination in the early phase and can be distinguished from those with $\tft\lesssim 5\times 10^{-4}$ through CV-limited CMB polarization observations. An accurate estimation of $\tft$, however, remains somewhat elusive. We investigate whether resampling the E-mode polarization data with limited spherical-harmonic modes may resolve this shortcoming.
\end{abstract}





\section{Introduction}
\label{sec:intro}
The epoch of cosmic reionization (EoR) marks an important era where the universe was lit up by astrophysical sources, and hydrogen atoms in the intergalactic medium (IGM) became progressively ionized. EoR begins with the birth of first stars and ends when the IGM becomes fully ionized. The IGM remains ionized after the end of EoR, due to the plethora of ionizing photons during the post-reionization epoch.

Cosmic reionization is not only sourced by astrophysical objects but also likely to have impacted the formation and evolution of relatively small-scale structures, and therefore probing the full history of cosmic reionization will be crucial in high-redshift astrophysics in general. For example, in terms of cosmological halos, minihalos (MHs) and low-mass atomic-cooling halos (LMACHs) are believed to have been subject to the Jeans-mass filtering of H II regions photo-heated to the temperature $T\gtrsim 10^{4}\,{\rm K}$ (e.g., \citealt{TW1996,Gnedin2000,Okamoto2008,HS2013}), while high-mass atomic-cooling halos (HMACHs) are blind to such a filtering due to their large virial temperature. 

Comic microwave background (CMB), especially its E-mode polarization anisotropy, bears information on the amount of free electrons in the universe through the Thompson scattering of CMB photons \citep[][and references therein]{Dodelson_book}. Nevertheless, with given observational limits, probing the full history of reionization used to be very difficult. Therefore, fitting CMB polarization measurements with two-parameter, i.e. the reionization optical depth $\tau$ and the duration of reionization  $\Delta z_{\rm reion}$, reionization models used to be sufficient. For this, tangent-hyperbolic form in $\xe(z)$ has been usually used (see e.g. \citealt{Planck2015}). CMB measurements have probed $\tau$ with increasing accuracy, to rule out reionization models with too large or too small $\tau$ \citep{wmap7,Planck2015,PlanckCollaboration2018}. As the accuracy of the CMB observation increased and the understanding of the foreground deepened, the best-fitting value to the data has shown a decreasing tendency: $\tau=0.088\pm 0.015$ \citep{wmap7}, $\tau=0.066\pm 0.016$ \citep{Planck2015}, $\tau=0.054\pm 0.007$ \citep{PlanckCollaboration2018}.

The polarization measurement by the Planck satellite seems to have reached a level that significantly constrains the history of reionization. In other words, parameterization of reionization with only two parameters, i.e. $\tau$ and the $\Delta z_{\rm reion}$, is now insufficient in fitting the CMB polarization anisotropy and biased toward a smaller value of $\tau$ than the actual value \citep{Mortonson2008, PlanckCollaboration2018}. Therefore, higher-order statistics for inferring reionization models became practical with the Planck results. The first method for such a higher-order statistics was the principal component analysis (PCA) based on the angular power spectrum of the  E-mode polarization anisotropy, $\ClEE$, proposed by \citet{Hu2003}. \citet{Mortonson2008} later showed that the two-component ($\tau$ and the $z_{\rm reion}$) analysis would bias the inferred $\tau$ to a smaller value than the true one that could be correctly inferred by the PCA. However, because the $E$-mode polarization depended rather weakly on the reionization history, some of the inferred reionization histories would contain non-physical values of $\xe$. A different type of higher-order statistics, sampling reionization histories directly with moving "knots" on the $z$--$\xe$ plane, was developed to avoid such non-physicality. They include methods by \citet{HS17} and \citet[][Flexknot]{Millea2018}. We will test all these methods in this paper.

In relation to the necessity for multiple parameters for reionization, an extended history of very early cosmic reionization due to strongly self-regulated Population III stars has been proposed \citep{Ahn2012,AS21} and tested against the CMB polarization data by the Planck \citep{Miranda,HS17,Heinrich2018,Millea2018,PlanckCollaboration2018,Qin2020_CMB,AS21,Heinrich2021,Wu2021}. In essence, such a model is characterized by a two-stage history: (1) the early, slow growth of $\xe$ for a wide range of redshift $z\sim$30--10 and (2) the late, rapid growth of $\xe$ for a short range of redshift $z\sim$10--6. Accordingly, the early stage of such a model can be tested by e.g. $\tft$, the CMB optical depth due to the free electrons at $z>15$.
\citet{Ahn2012} simulated cosmic reionization numerically in a box large enough ($\sim 150 \,{\rm Mpc}$) to provide the statistical significance but at the same time encompassing the full dynamic range of cosmological halos from MHs to HMACHs under varying degrees of feedback effects, overcoming the usual numerical limit in resolving halos encountered by large-box simulations. 
\citet[][AS21 hereafter]{AS21} inherited the gist of the model studied by \citet{Ahn2012} but explored a wider parameter space through a semi-analytical calculation. With a reasonable degree of radiative feedback effect, parameterized by the threshold Lyman--Werner radiation intensity $J_{\rm LW,th}$, Population III stars inside MHs are found to provide an "extended high-redshift ionization tail" history at $z\gtrsim 10$ through self-regulation (\citealp{Ahn2012}; AS21). This unique feature had indeed been found to be preferred by the Planck 2015 data \citep{Miranda,Heinrich2018}, when the observation of the CMB polarization anisotropy was made through the Low-Frequency Instrument (LFI). Later, after the addition of the High-Frequency Instrument (HFI) data, \citet{Millea2018} and \citet{PlanckCollaboration2018} showed that $\tft$ was likely to be much smaller than values inferred by \citet{Miranda} and \citet{Heinrich2018}.

The Bayesian inference of $\tft$ has been somewhat eventful.  The estimated $\tft$ converged to a common value only recently, after some confusion triggered by the original estimation by \citet{PlanckCollaboration2018}. The possibility for many pronounced high-redshift ionization scenarios had been strongly limited by the initial claim by the Planck Collaboration, showing that the Planck Legacy Data (PLD) gives $\tft<0.007$ at $2\sigma$ level without a lower bound \citep{PlanckCollaboration2018}. For this claim, \citet{Millea2018} and \citet{PlanckCollaboration2018} sampled reionization histories directly from the $z$--$\xe$ plane with a method called FlexKnot analysis and performed a Bayesian inference. However, AS21 found that a model with $\tft=0.008$, which is beyond the $2\sigma$ limit set by \citet{PlanckCollaboration2018}, is indeed among the models best-fitting PLD. This finding was then backed up by the re-analysis of the PLD by the principal component analysis (PCA; \citealp{Heinrich2021}) and a simple likelihood analysis \citep{Wu2021}, allowing more pronounced high-redshift ionization tails by yielding a boosted upper limit $\tft<0.020$ at $2\sigma$ level. Now roughly in agreement with the latter results, the Planck Collaboration has recently corrected their original claim on $\tft$ to $\tft<0.018$ at $2\sigma$ level \citep{Planck_erratum}. Therefore, currently it seems that the extended, high-redshift ionization by Pop III stars is still a viable scenario. Even though \citet{Wu2021} claimed that probing $\tft$ with even CV-limited CMB observation would be impractical mainly due to the smallness of $\tau$, we will prove otherwise in this paper.

Advances in the ground and space CMB telescopes seem very promising. 
High-precision CMB apparatuses will be operating in the near future that might enable probing e.g. $\tft$ soon.
While recent CMB polarization measurement on large angular scales by Planck are limited by instrumental noise, near-future CMB polarization experiments such as CLASS \citep{2021arXiv210708022D}, Groundbird \citep{2021ApJ...915...88L}, and PIPER \citep{2016SPIE.9914E..1JG}, will measure the E-mode polarization anisotropies on large angular scales limited only to the sample variance: namely, the precision is limited not due to the instrumental noise but to the limited number of samples of the Fourier modes we can observe. In particular, the LiteBIRD satellite \citep{2020SPIE11443E..2FH}, to be launched in the late 20's, will perform polarimetric observation over the entire sky with fifteen frequency bands to mitigate the foreground problem.
Although the main aim of the LiteBIRD experiment is to measure the B-mode angular power spectrum to detect the inflationary gravitational waves, a cosmic variance (CV) limited E-mode angular power spectrum on large scales, which is beneficial to the EoR study, will be an important and guaranteed product from the experiment.


Can we probe the history of reionization better in the future than with existing CMB observations, and if so, how well? This is the question we try to answer in this paper. Motivated by the theoretical and observational developments, we study how much a CV limited apparatus could constrain the history of reionization, especially the high-redshift ($z>15$) part. Toward this end, we first adopt the theoretical reionization models  by AS21 to generate several mock CMB data. We then perform a Bayesian inference through the Markov-Chain Monte-Carlo (MCMC) sampling of reionization histories with various sampling methods to estimate cosmological parameters and the base reionization history. A key point we address will be whether one could probe $\tft$ even when $\tau$ is as small as $\tau\sim$0.055--0.060, which was a part of reasons for \citet{Wu2021} to present a pessimistic view on the possibility for probing $\tft$. In essence, the main goal of our paper is the same as that by \citet{Watts_forecast}, who have performed Fisher-matrix estimation on $\tau$ based on composite tangent-hyperbolic reionization models \citep{Heinrich2021}. With our approach described above, we will not be limited to Gaussianity that in general limits the reliability of Fisher matrix analysis, and will also try to find which sampling method would be most optimal for probing the history of reionization.

The paper is organized as follows. In Section 2, we describe the process for the generation of mock CMB data, various methods for sampling reionization histories, and the likelihood of a sample in a few variants against the mock CMB. In Section 3, we present out main results on the inferred reionization-related parameters and constraints on reionization histories. In section 4, we summarize the result and discuss theoretical and observational prospects.

\section{Methodology}
\label{sec:methodology}
\subsection{Mock CMB data: base reionization models}
\label{subsec:base_reion_model}

We assume that LiteBIRD, or some similar apparatus whose CMB polarization data is only CV-limited, measures a sky whose true reionization history is given by models in AS21. AS21 have categorized reionization models for different feedback effects on the formation of radiation sources: (1) vanilla (V) models sourced by atomic-cooling halos (ACHs) without any feedback, (2) self-regulated type I (SRI) models sourced by ACHs but with low-mass atomic-cooling halos (LMACHS) modulated by the Jeans-mass filtering inside H II regions, and (3) self-regulated type II (SRII) models sourced by both ACHs and minihalos (MHs), but LMACHS modulated by the Jeans-mass filtering inside H II regions and MHs modulated by the Lyman-Werner radiation feedback. All model categories can generate reionization histories with or without substantial high-redshift ionization tail (i.e. ionization beyond $z\gtrsim 15$). However, models satisfying both $\tau=0.054 \pm 0.007$ and $z_{\rm end}\sim 5.5$, as constrained by the Planck \citep{PlanckCollaboration2018} and the Lyman-$\alpha$ forest observations (e.g. see \citealt{Qin_zend} and references therein) respectively, form substantially limited sets from each category. Under these restrictions, for example, vanilla models are not allowed to have any significant high-redshift ionization tail and  $\tft\lesssim 5\times10^{-4}$, while SRII models can have significant high-redshift ionization tails with $\tft \sim 8\times10^{-3}$. A natural question is whether CV-limited CMB experiments could discriminate these models and break the degeneracy in $\tau$.

\begin{figure}
    \centering
    \includegraphics[width=8cm]{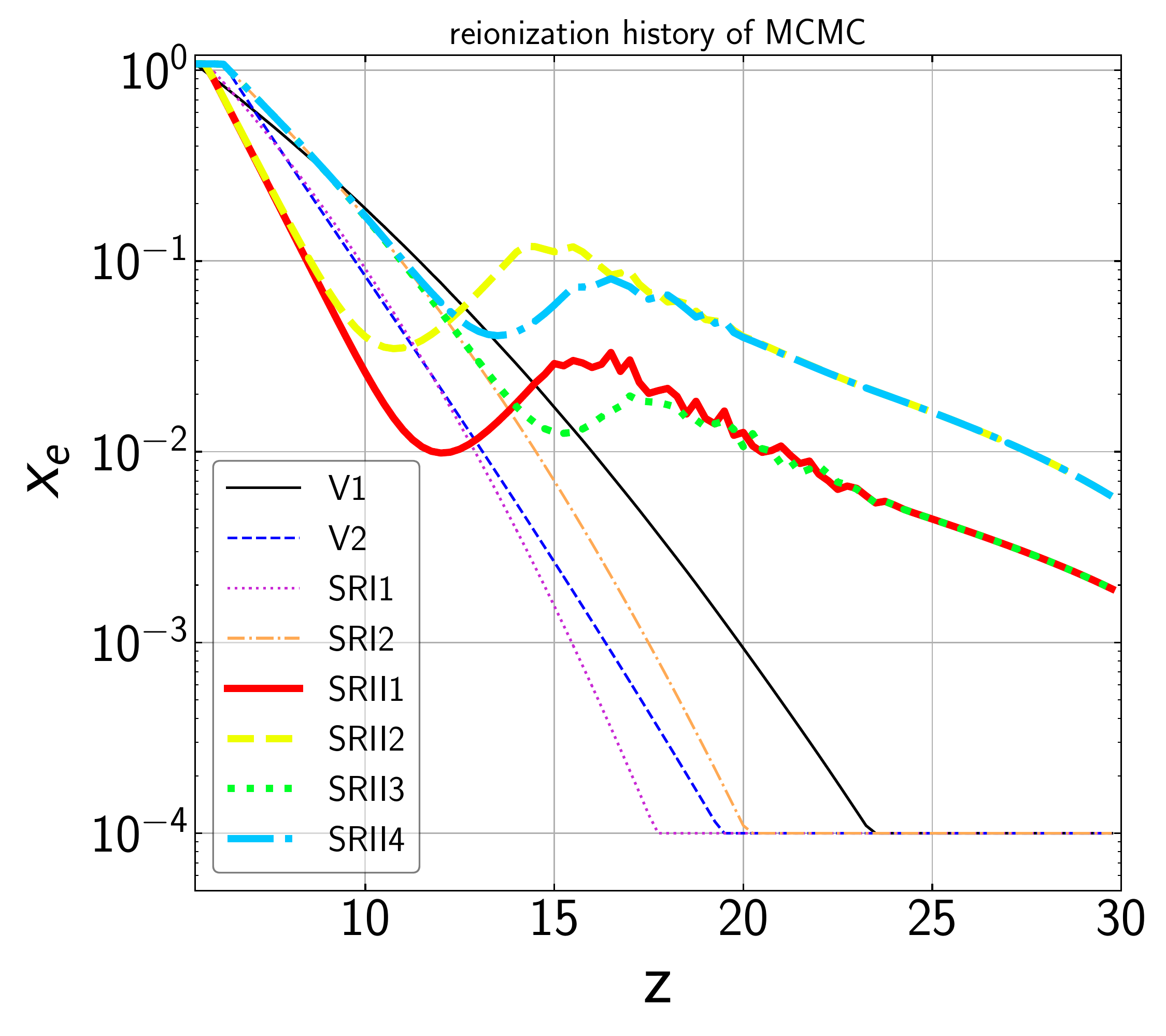}
    \caption{Time evolutions of the ionization fraction $x_e$ in our reionization models.}
    \label{fig:IH}
\end{figure}

From AS21, we select several models that can fit the E-mode polarization power spectrum of the Planck well (Table \ref{tab:models} and Figure \ref{fig:IH}; with one exception as described below). These will be used to test the discriminating power of various Bayesian inference methods, that are described in Sections \ref{subsec:PCA} -- \ref{subsec:HS17}. We select two representative models to generate mock CMB data: V1 and SRII2. These two models have similar $\tau$'s but with very different reionization histories. Model V1 has $\tft=4.0\times 10^{-4}$ while model SRII2 has $\tft=7.8\times 10^{-3}$ and can reach $\xe \sim 0.1$ at $z=15$ thanks to Pop III stars inside MHs (see AS21 for details). In order to generate mock CMB data, we use a version of CAMB that is modified \citep{Mortonson2008} to incorporate general forms of $\xe$ that are not restricted to the usually used tangent-hyperbolic from. 

Note that the base model histories quantified by $\xe (z)$ are generated under a fixed set of cosmological parameters that fit the Planck Legacy Data best \citep{PlanckCollaboration2018}. In principle, cosmological parameters that impact the structure formation will change the history of reionization for given astrophysical parameters. For example, changes in the present matter content $\Omega_{m,0}$ will change the resulting reionization history for the given ionizing-photon escape fraction $f_{\rm esc}$, star formation efficiency $f_*$, and number of ionizing photons per stellar baryon $N_{\rm ion}$ in the vanilla model of AS21. In this work, we do not intend to constrain such astrophysical parameters and simply take these fixed $\xe (z)$'s for estimating only the reionization history itself and its derived parameters such as $\tau$ and $\tft$.

\begin{table}
    \caption{Reionization models used in the analysis, with the total
      optical depth, $\tau$, and the fractional optical depth at
      $z>15$, $\tau_{z>15}$. Model names are from AS21, and renamed
      aliases are used in this work. Two representative cases with
      (SRII2) and without (V1) a high-redshift ionization tail are
      used to generate mock CMB data.}
    \centering
    \begin{tabular}{l|c|c}
         model name (renamed aliases) &  $\tau$ & $\tft$ \\
         \hline
         V-L\_dF (V1) & 0.05998 & 3.97E-4 \\
         V-M1\_F (V2) & 0.05555 & 4.71E-5 \\
         SRI-L0\_dF (SRI1) & 0.05436 & 2.01E-5 \\
         SRI-LL\_F (SRI2) & 0.06140 & 1.16E-4 \\
         SRII-L0-300-e0.5-J0.05 (SRII1) & 0.05063 & 2.29E-3 \\
         SRII-L0-300-e1.0-J0.10 (SRII2) & 0.05960 & 7.79E-3 \\
         SRII-LL-300-e0.5-J0.05 (SRII3) & 0.06330 & 1.83E-3 \\
         SRII-LL-300-e1.0-J0.10 (SRII4) & 0.06944 & 6.96E-3
    \end{tabular}
    \label{tab:models}
\end{table}

\subsection{Sampling reionization histories}
\label{subsec:sampling_reion}
\subsubsection{Principal Component Analysis}
\label{subsec:PCA}
The principal component analysis (PCA) of the reionization history of the universe using the CMB E-mode polarization anisotropies was first studied by \cite{Hu2003}. Here we briefly review the PCA analysis and introduce the principal component on linear and log-space basses.

\begin{figure}
    \centering
    \includegraphics[width=8cm]{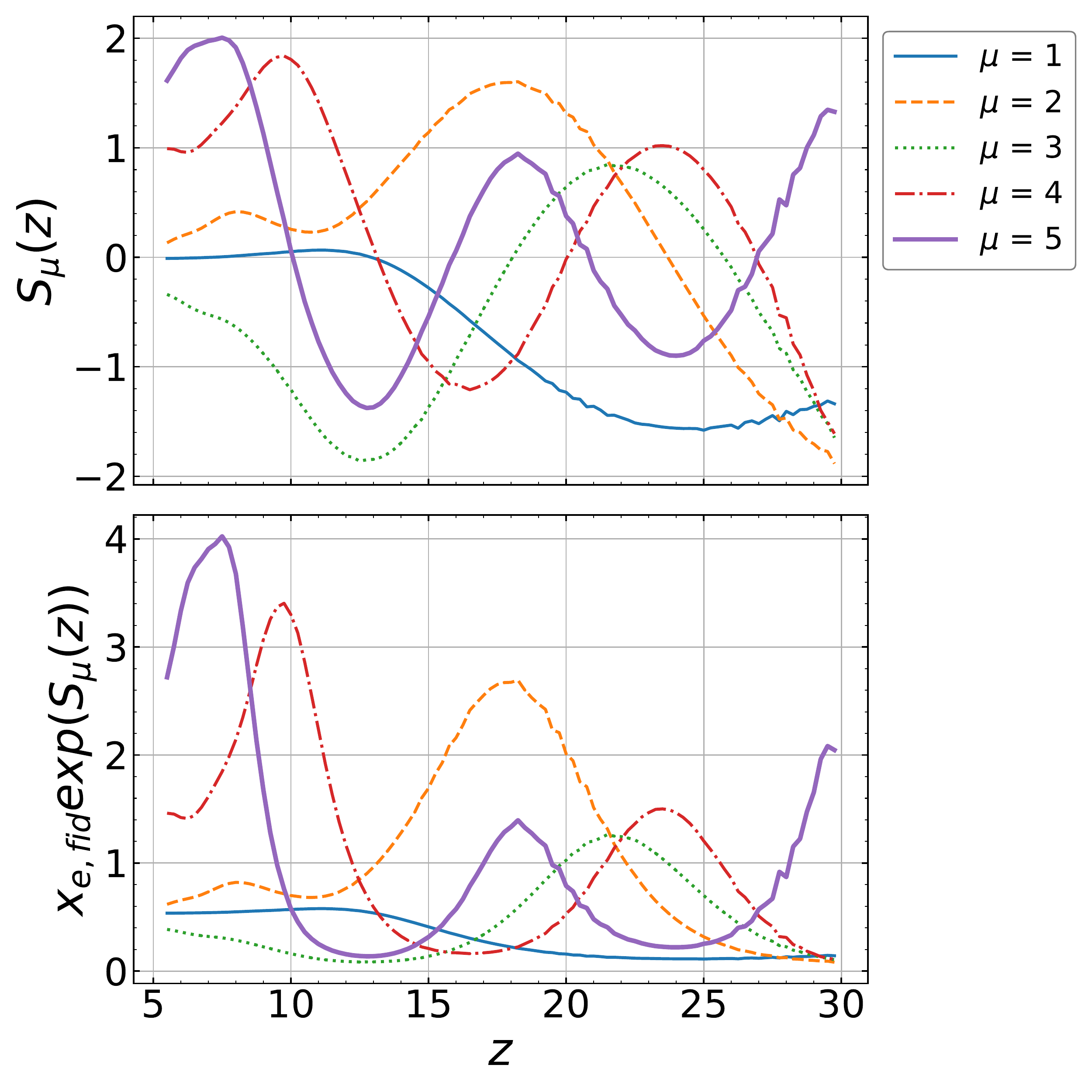}
    \caption{First five eigenmodes of the PCA analysis on the linear (top) and log (bottom) bases. As \citet{Mortonson2008} showed that, since the larger $\mu$ corresponds to the larger frequency mode, the effect on $\ClEE$ becomes smaller as $\mu$ becomes larger. Therefore the mode of $\mu =1$ contributes dominantly to the total optical depth.}
    \label{Eigen}
\end{figure}

On the linear basis, we parameterize the time evolution of the ionization fraction of the universe as
\begin{eqnarray}
    \xe(z) =\xefid(z)+{ \sum^{N_\mathrm{PC}}_{\mu=1} m_{\mu}S_{\mu}(z) }
    \label{eq:eq1}
\end{eqnarray}
where $S_\mu(z)$ are the principal component (PC) eigenmodes, and $m_\nu$ are the PC components. The fiducial reionization model $\xefid$ is taken as {$\xefid=0.5(1+f_{\rm He})$} to minimize the probability to realize unphysical reionization models ($x_e<0$) \cite{Millea2018}. Because the PC eigenmodes are mutually independent in the given redshift range of $[z_{\rm min},z_{\rm max}]$, the PCs are calculated from $x_e(z)$ as
\begin{eqnarray}
    m_{\mu} =\frac{\int ^{z_\mathrm{max}}_{z_\mathrm{min}}dz S_\mu(z)  \left( \xe(z)- \xefid(z)\right) }{z_\mathrm{max}-z_\mathrm{min}}~.
\end{eqnarray}
The eigenmodes $S_\mu(z)$ are obtained based on the Fisher information matrix for $x_e(z)$ in the redshift range from cosmic variance E-mode measurements as
\begin{eqnarray}
F_{ij}&\equiv& \sum^{\ell_\mathrm{max}}_{\ell=2}\biggl(\ell+\frac{1}{2} \biggr) \frac{\partial \log \ClEE}{\partial x_\mathrm{e,fid}(z_i)}
    \frac{\partial \log \ClEE}{\partial x_\mathrm{e,fid}(z_j)}\nonumber \\
    &=& (N_z+1)^{-2}\sum^{N_{\rm PC}}_{\mu=1}~S_\mu(z_i) \lambda_{\mu}S_\mu (z_j)
\label{eq:eigen}
\end{eqnarray}
where $N_{\rm PC}$ is the number of PC components, $\lambda_\mu$ are the eigenvalues, and $N_z$ is the number of bins in redshift. Following the discussion in \cite{Hu2003}, we take $N_{\rm PC}=6$, $z_{\rm min}=5.5$ and $z_{\rm max} = 30$, and discretize the redshift range with $\Delta z = 0.25$. 
The eigenmodes $S_\mu$ can be normalized as 
\begin{eqnarray}
    \sum^{N_{\rm PC}}_{\mu=1}~S_\mu(z_i)S_\mu (z_j)=(N_z+1)\delta_{ij}~.
\end{eqnarray}
We determine the eigenvalue $\lambda_\mu$ in descending order when diagonalizing the Fisher-matrix.
In the top panel of Figure~\ref{Eigen}, we depict the eigenmodes of the PCA analysis obtained from Eq.~(\ref{eq:eigen}).

In addition to the standard PCA analysis described above, we parameterize the reionization history using the PCA technique in log space, aiming to avoid the non-physical reionization histories that have negative ionization fractions at some redshifts. The Fisher matrix and the eigenmodes are defined by
\begin{eqnarray}
F_{ij}&=& \sum^{\ell_\mathrm{max}}_{\ell=2}\biggl(\ell+\frac{1}{2} \biggr) x_{e, {\rm fid}}(z_i)\frac{\partial \log \ClEE}{\partial x_\mathrm{e,fid}(z_i)}
    x_{e,{\rm fid}}(z_j)\frac{\partial \log \ClEE}{\partial x_\mathrm{e,fid}(z_j)}\nonumber \\
    &=& (N_z+1)^{-2}\sum^{N_{\rm PC}}_{\mu=1}~\tilde{S}_\mu(z_i) \lambda_{\mu}\tilde{S}_\mu (z_j)
\end{eqnarray}
where $\tilde{S}$ is the eigenvector in the log space basis, and we take $\xefid(z)$ to be the constant that is the same as in the linear PCA analysis. Thus the Fisher matrix is also the same as that in the linear PCA analysis up to a multiplicative constant. The reionization history in the log space basis is expressed as
\begin{eqnarray}
    \xe(z) &=&\xefid \exp{\left( \sum^{N_\mathrm{PC}}_{\mu=1} m_{\mu}\tilde{S}_{\mu} \right)}~.
\end{eqnarray}
From the above expression, it is evident that the ionization fraction is positive definite. We note, however, that we should impose a physicality condition on the constraints on $m_\mu$ since the ionization fraction tends to be unphysically large ($x_e\gg 1$) in the PCA analysis in log space.

\subsubsection{FlexKnot Analysis}
\label{subsec:FlexKnot}
FlexKnot analysis was used by MB to take a prior different from the PCA and also to  guarantee the physicality of $x_e$ and to directly infer $x_{e}(z)$. FlexKnot samples $\{z,\,x_e\}$ points uniformly with a limited number of "knots" and then interpolate the knots at uniformly sampled redshifts to sample one reionization history. This method, in cosmology, was first used to probe any feature differing from the pure power law in the primordial power spectrum when analyzing the CMB anisotropy \citep{Vazquez2012,Planck_flexknot}. MB also tried forcing a uniform prior on $\tau$ and found that the posterior of $\tau$ was affected, even though one can question whether such a prior is justified for a derived parameter such as $\tau$.

We apply the FlexKnot analysis to the mock LiteBIRD data, but with two different choices. First, we use the original FlexKnot analysis with the condition$z^\mathrm{ knot}_{1}<z^\mathrm{ knot}_{2}< \dotsb<z^\mathrm{knot}_{N_\mathrm{ knot}}$.
This condition was enforced in the original version of FlexKnot in inferring the primordial power spectrum \citealt{Planck_flexknot}, where of course the wave number $k$ takes place of $z$. Second, we also use a slightly modified version of FlexKnot, which we denote as FlexKnot-m (modified). In FlexKnot-m, we divide the full EoR domain $z\otimes x_{e}=[6,\, 30]\otimes[0,\,1+f_{\rm He}]$ into $N_{\rm knot}$ rectangular sub-domains when using $N_{\rm knot}$ knots. The sub-domains are sequential and the $i$th sub-domain has $z\otimes x_{e}=[z_{i,L},\, z_{i,R}]\otimes[0,\,1+f_{\rm He}]$. Then, each knot is sampled uniformly in each of these sub-domains, and obviously the same condition $z^{\rm knot}_{1}<z^{\rm knot}_{2}<\dotsb<z^{\rm knot}_{N_{\rm knot}}$ is met. Note that the FlexKnot-m takes a prior different from that of the FlexKnot by limiting the domain accessible to each knot.

We set $N_{\rm knot}=4$ in this work. Even though a complete analysis should involve varying $N_{\rm knot}$ and weighting each posterior with the evidence, we just take this approach for simplicity and defer such a complete analysis to the future. In both FlexKnot and FlexKnot-m, we use the piecewise cubic Hermite interpolating polynomial (PCHIP) scheme for interpolation, which guarantees the physicality of $x_e$ as long as $x^{\rm knot}_{e,\,i}>0$ for any $i$. This is because PCHIP guarantees monotonic change of a derived function $y(x)$ in any given piecewise interval $[x_{i}, x_{i+1}]$ where $x_{i}$ and $x_{i+1}$ are adjacent points in $x$.



\subsubsection{HS17 Analysis}
\label{subsec:HS17}
The HS17 analysis by \citet{HS17} is, in some sense, a restricted version of the FlexKnot analysis. While knots are used in the HS17 analysis, these knots are located at fixed redshifts $z^{\rm knot}_{i}$ and only $\xe$'s at each $z^{\rm knot}_{i}$ are varied. Once knots are defined, these knots are interpolated at uniformly-spaced redshifts to sample one reionization history.

We use the HS17 analysis here, to examine how its predictive power compares to PCA and FlexKnot. Another motivation is to investigate whether one can find any scheme to probe the the ``reionization dip'' (or $\xe$ showing a non-monotonic behavior in time) caused by a global recombination at $z\sim 10 \text{--} 12$, which can be observed in most SRII models (Figure \ref{fig:IH}). If this were to affect the CMB polarization anisotropy to any extent, such a feature may require at least three knots, and possibly be better fitted by a larger number of knots. Therefore, we either choose three or seven knots in the HS17 analysis, which we will denote as HS17-3 and HS17-7, respectively.

The prior of HS17 is obviously different from other analyses and is not uniform on the $z$--$\xe$ plane. However, we believe that this should not be a reason to discard the analysis. We simply take a route to test various methods and see which method is the most optimal one in finding out the  fiducial model for the mock CMB data.
%
%


\subsection{CMB likelihood analysis}
\label{subsec:likelihood}
\subsubsection{mock CMB likelihood}
\label{subsec:usual_likelihood}
For an evaluation of the CMB likelihood of the reionization models described by the PCA, Flexknot, and HS17 schemes, we assume a simple exact chi-squared likelihood. Specifically, given the observed (mock) T and E-mode angular power spectra, $\hat{C}_\ell^{TT}$,$\hat{C}_\ell^{TE}$, and $\hat{C}_\ell^{EE}$, we calculate the chi-squared given by (see, e.g., \citealt{2006JCAP...10..013P})
\begin{equation}
\chi^2_{\rm eff} = -2\ln{\cal L} = \sum_{\ell=2}^{\ell_{\rm max}} (2\ell+1) f_{\rm sky}
\left(
\frac{D}{C}+\ln \frac{C}{\hat{C}}-2
\right)~,
\label{eq:chi2}
\end{equation}
where $D$, $C$ are defined as
\begin{eqnarray}
&&D(C_\ell,\hat{C}_\ell)= \hat{C}_\ell^{TT} \ClEE-2 \hat{C}_{\ell}^{TE} \ClTE +  \ClEE \ClTT~, \\
&&C(C_\ell) = \ClTT \ClEE-(\ClTE)^2~, 
\end{eqnarray}
and $\hat{C}=C(\hat{C}_\ell)$. Here $C_\ell$ without hat denotes the
model power spectrum given by
$C_\ell = C_\ell^{\rm theory}+N_\ell^{\rm noise}$, where $C_\ell^{\rm theory}$ is the
theoretical power spectrum and $N_\ell$ is the power spectrum of the
assumed instrumental noise.

\begin{figure}
    \centering
    \includegraphics[width=8cm]{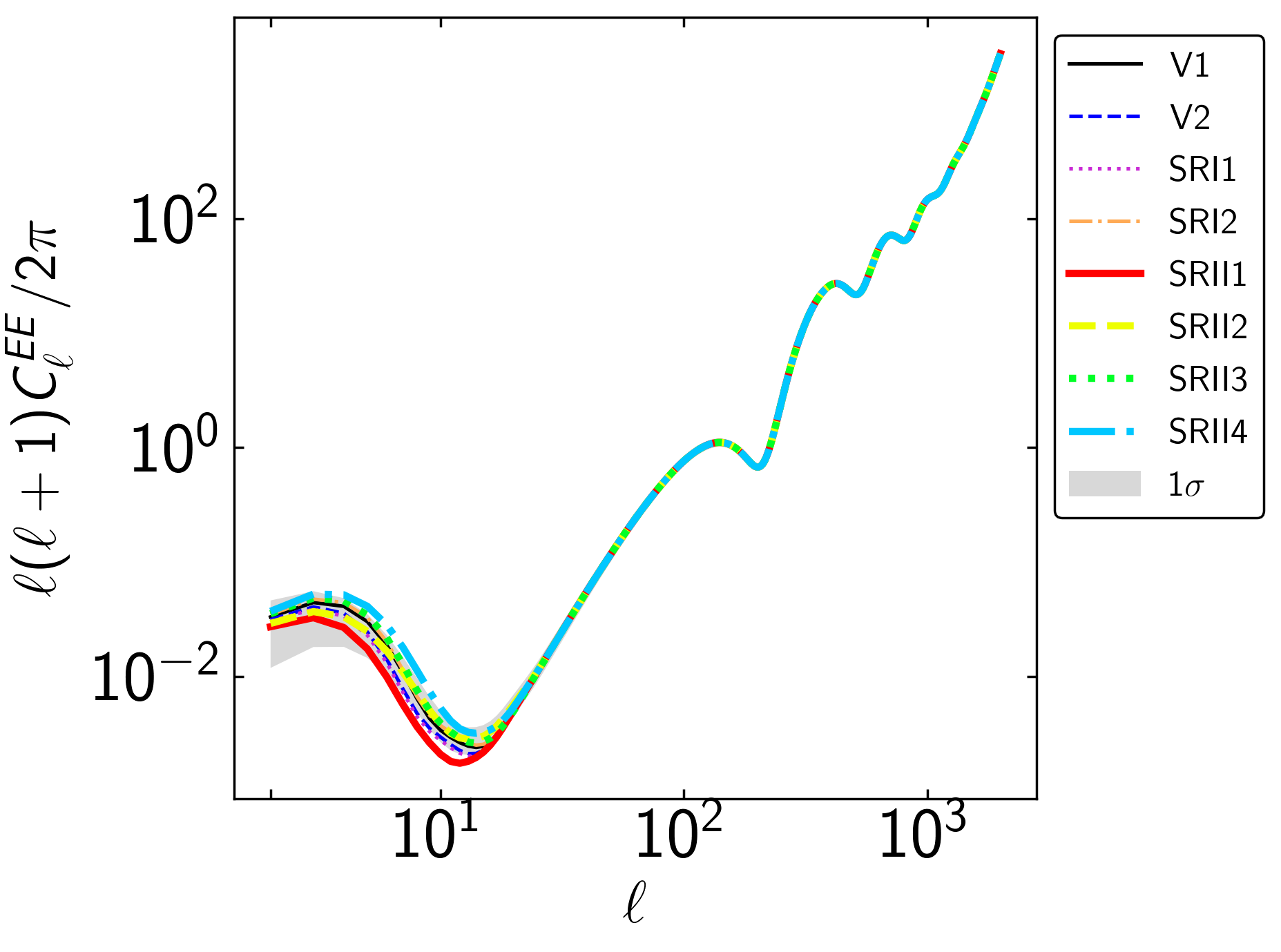}
    \includegraphics[width=8cm]{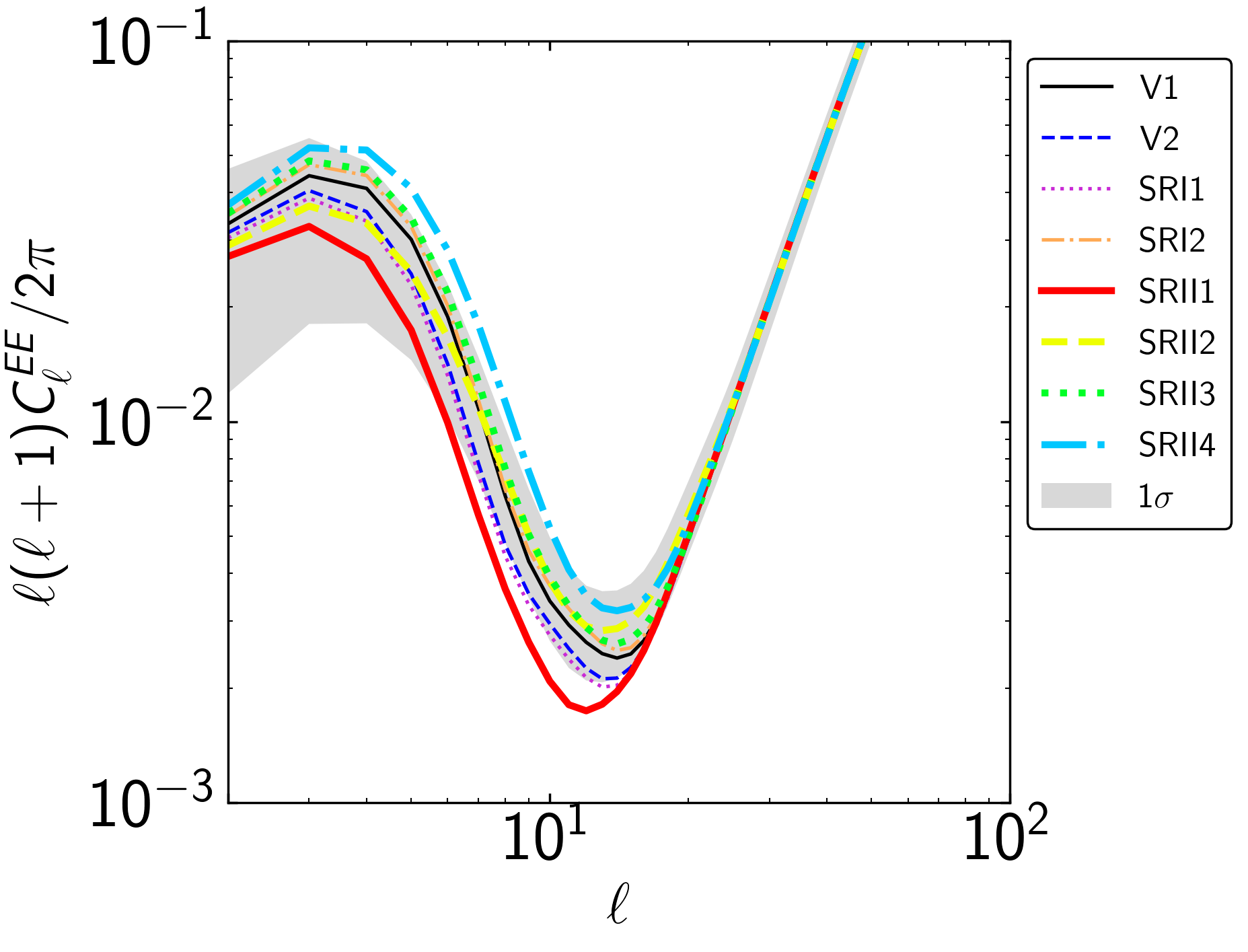}
    \caption{Power spectra of CMB E-mode polarization $\ClEE$. Lines are theoretical power spectra including noise contribution expected for LiteBIRD-like experiments. The error bars show only the size of the cosmic variance.}
    \label{m1tom2_L0Cl_2}
\end{figure}

For the instrumental noise we assume
$\sigma_{\rm pol} = 2$~$\mu$Karcmin at $\ell \le 1000$, which is a
rough estimation of the noise of LiteBIRD. This noise is small enough
to make LiteBIRD a practically CV-limited apparatus. The fiducial
E-mode polarization power spectrum and the size of the assumed
errors\footnote{We made
the error bars in the following steps. First, we wrote the $\chi^2$
distributions with mean $<C_\ell+N_\ell>$ and degrees of freedom
(2$\ell$+1) for each $\ell$. Second, we plotted the region of this
distribution excluding 16\% of both ends on
Figure~\ref{m1tom2_L0Cl_2}. This operation resulted in a 1$\sigma$
confidence region.}, based on SRII2, are 
depicted in Figure~\ref{m1tom2_L0Cl_2} against models listed in
Table~\ref{tab:models}.

\subsubsection{CMB likelihood for optimally probing high-$z$ optical depth}
\label{subsec:tau15}
$\tft$ is a parameter that is useful in quantifying how active the reionization process was in its early phase, or more explicitly during $z>15$. The impact of the high-redshift reionization on $\ClEE$, however, may make it not the most optimal way to use the CMB polarization data in the full range of $\ell$. As observed in \citet[Figure 8]{AS21}, the information on $\xe(z>15)$ seems strongly encoded in  $\ClEE$ in a limited range of $\ell$, or $15\lesssim \ell\lesssim 25$, while $\ClEE$'s in the low-$l$ regime ($\ell\lesssim 14$) seem less strongly coupled to $\xe(z>15)$ than $\ClEE(15\le \ell\le25)$. For $\ell \gtrsim 30$,  $\ClEE$ is determined mostly by the physics during the pre-recombination epoch \citep{Watts_forecast}, which is beyond the interest of this paper.

We thus try to find an optimal way of probing $\tft$ by using the CMB likelihood with limited $\ell$ values. First, one may try $\chi^2_{\rm eff}$ in the following form:
\begin{equation}
\chi^2_{\rm eff}(15\le \ell\le25) \equiv \sum_{\ell=15}^{25} (2\ell+1) f_{\rm sky}
\left(
\frac{D}{C}+\ln \frac{C}{\hat{C}}-2
\right)~.
\label{chi1525}
\end{equation}
Second, one may instead use the following form:
\begin{equation}
\chi^2_{\rm eff,\,E}(15\le \ell\le25) \equiv \sum_{\ell=15}^{25} (2\ell+1) f_{\rm sky}
\left(
\frac{\hat{C}_{\ell}^{EE}}{\ClEE}+\ln \frac{\ClEE}{\hat{C}_{\ell}^{EE}}-1
\right)~,
\label{chi1525Eonly}
\end{equation}
keeping only the E-mode angular power spectrum. Finally, it is also possible to use the following $\tau$-modulated form:
\begin{equation}
\tilde{\chi}_{\rm eff,\,E}^{2}(15\leq \ell\leq 25) = \chi^2_{\rm eff,\,E}(15\leq \ell\leq 25) + (\tau-\bar\tau)^2/\sigma_\tau^2 , 
\label{chi1525Eonlytauprior}
\end{equation}
where $\sigma_\tau$ is the standard deviation of $\tau$ in the posterior distribution from the full $TT$+$TE$+$EE$ analysis without limiting $\ell$. \footnote{The multipoles at $15<\ell<25$ should be removed to avoid double counting of statistical information. However, we checked that the constraints on the total optical depth do not change much with or without removing these multipoles.}

Limiting $\ell$ in equations (\ref{chi1525}) and (\ref{chi1525Eonly}) is hinted from \citet{AS21} as described above, and also partly from \citet{Watts_forecast} who showed that the precision in the estimated $\tau$ depends on the range of $\ell$ chosen. We try equation (\ref{chi1525Eonly}) in addition to equation (\ref{chi1525}), to check whether the impact of the high-redshift reionization is better probed only by the E-mode analysis in such a restricted range of $\ell$ and thus inclusion of $TT$ and $TE$ may be a nuisance. Because main cosmological parameters cannot be inferred only with the $E$-mode, we fix all the cosmological parameters when using equation (\ref{chi1525Eonly}).

The motivation to use equation (\ref{chi1525Eonlytauprior}) is as follows. If one were to use a limited-$\ell$ scheme to estimate $\tft$ without the prior knowledge on $\tau$, there exists a danger of wrongfully estimating $\tau$ due to the limited information and thus the estimation of $\tft$ can also be affected. For example, from the mock data based on SRII2, FlexKnot yields $\tau=0.0634\pm 0.0017$ and the posterior of $\tau$ is usually very close to gaussian. This is then fed into the resampling analysis in the form of equation (\ref{chi1525Eonlytauprior}).

\section{Results}
\label{section:results}
\subsection{PCA analysis}
\label{sub:PCAresult}
\begin{figure}
    \centering
    \includegraphics[width=8cm]{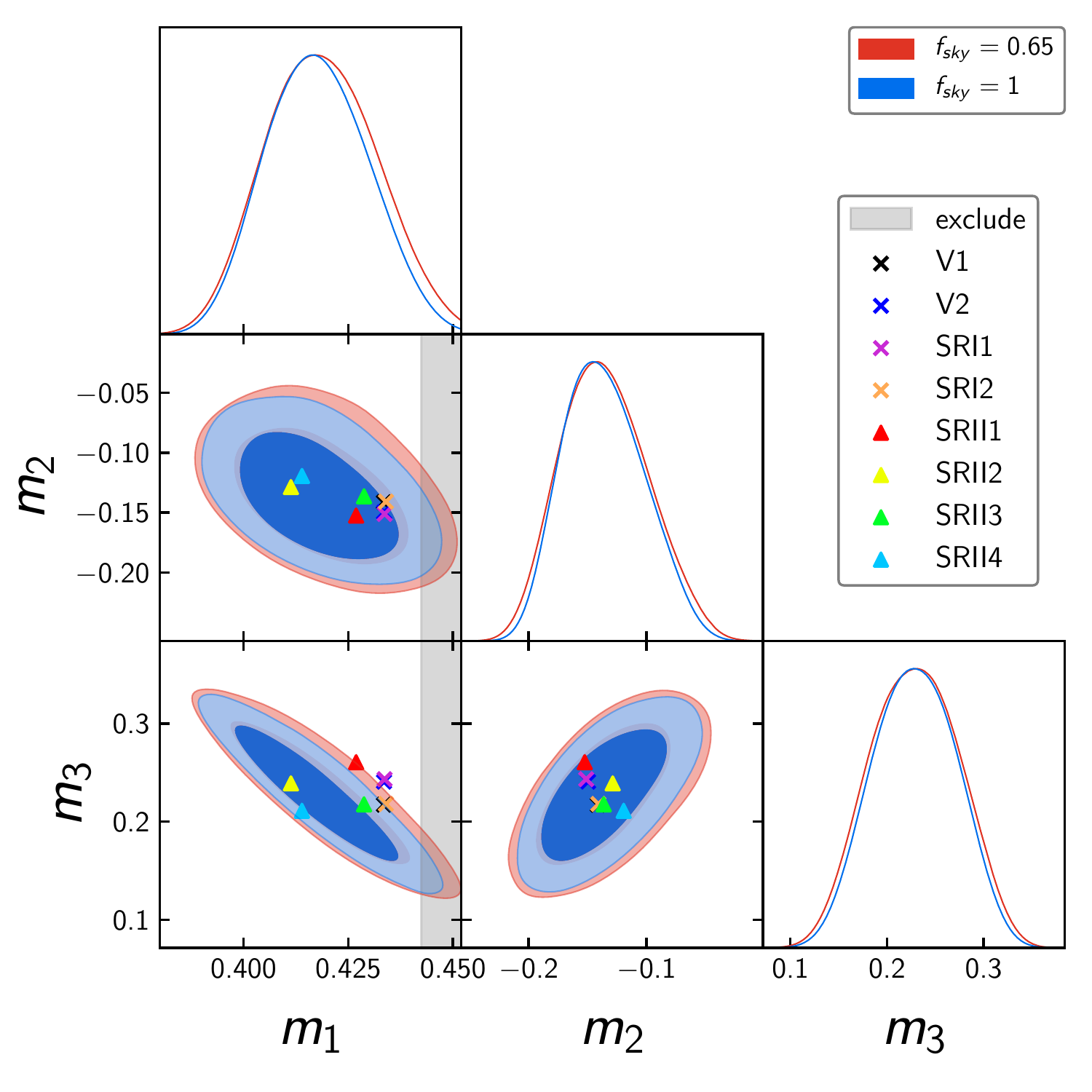}
    \caption{Constraints (1$\sigma$ and 2$\sigma$ confidence levels) on the PCs in the linear PCA analysis.
    The mock data is based on SRII2 model.
    The points represents the PCs of the reionization models described in the figure legend. The gray shaded regions are where $x_e$ violates the physicality condition $x_e<0$. Cases with $f_{\rm sky}=0.65$ and $f_{\rm sky}=1$ are represented in red and blue shades, respectively.}
    \label{fig:tr}

    \includegraphics[width=8cm]{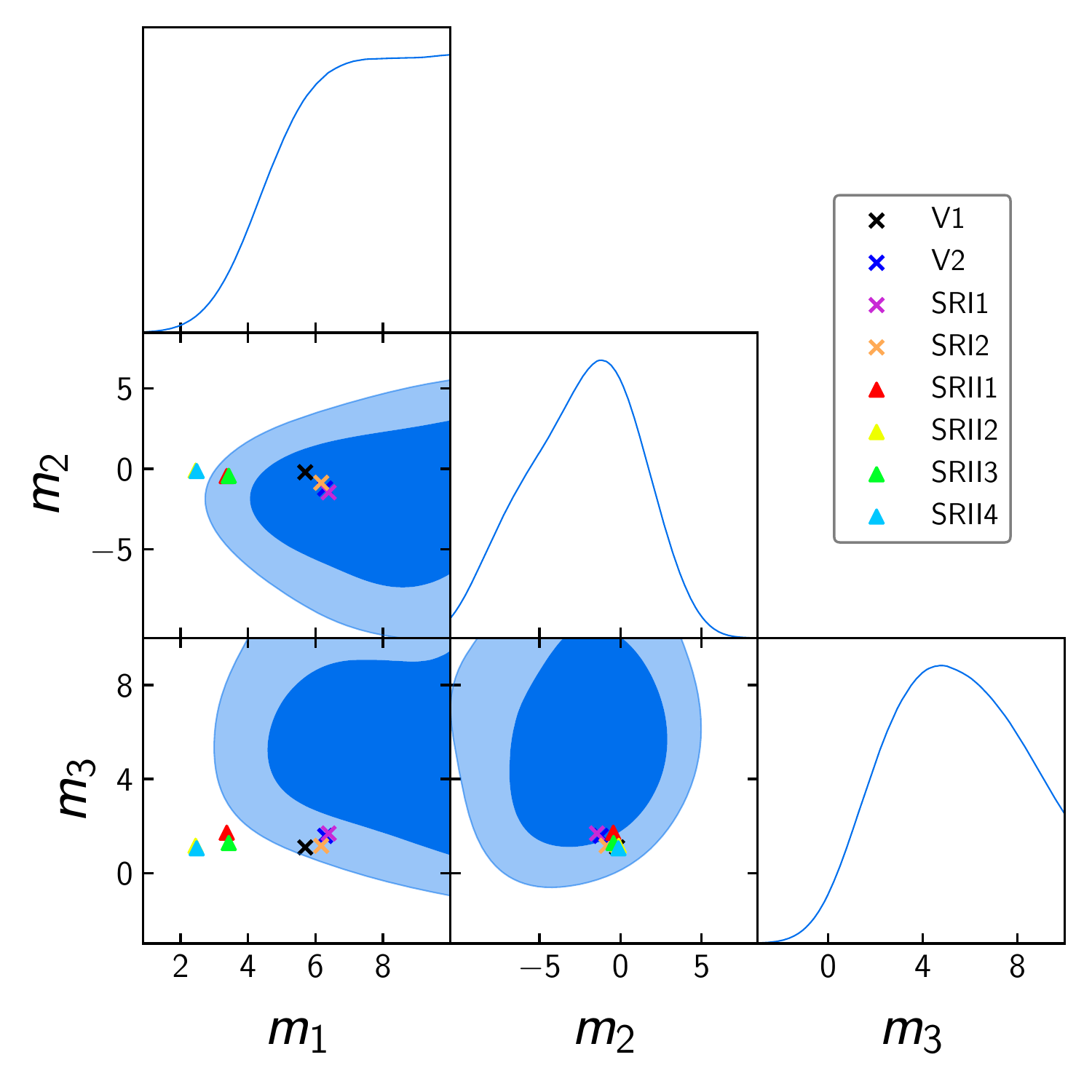}
    \caption{same as figure\ref{fig:tr}, but using the logarithmic basis. This is for $f_{\rm sky}=0.65$.}
    \label{fig:contours_log_PCA}
\end{figure}

We perform MCMC analyses varying the PCA parameters $m_\mu$ $(\mu=1,\cdots 6)$ and the standard six cosmological parameters. In the linear PCA analysis, we found that the third PCA parameter, $m_3$, is most sensitive to the contribution from the minihalos. In Figure~\ref{fig:tr}, we show the triangle plots for the posterior distributions for the PCA parameters $m_1$, $m_2$ and $m_3$. The fiducial model we chose is the SRII2 model and its PCs are roughly placed in the center of the $1\sigma$ regions. Note that, in the linear PCA basis, some models are indistinguishable in the $m_1$, $m_2$ and $m_3$ parameter space. In the left-bottom panel, we find that the models, V2 and SRI1, are outside the $2\sigma$ contour.  These models are representative models that do not contain early reionization contributions.

The model-discrimination power of CV-limited experiments, with PCA, lies mostly in the 2-dimensional parameter space for two major principal components, and it is indeed quite impressive. Models with high-redshift ionization tails (SRII2 and SRII4) are distinguished from tail-less models (V1, V2, SRI1 and SRI2) at $\sim 2\sigma$ level. Even in the same category SRII, weakly tailed models (SRII1 and SRII3) may be distinguished from strongly tailed models (SRII2 and SRII4) at $\sim 1\sigma$. This is true for $f_{\rm sky}=0.65$ as seen in Figure~\ref{fig:tr}, and is better for $f_{\rm sky}=1$. Note that the difference in the total optical depth between e.g. SRII2 and SRI2 is $\Delta \tau = 0.0018$, which is less than the full-sky cosmic variance error of $\sigma_\tau = 0.002$ \citep{Millea2018}.

In Figure~\ref{tau_tougou} and Figure~\ref{tau15_tougou}, we show the marginalized distributions of the optical depth, $\tau$, and the fractional optical depth at $z>15$, $\tft$, in the V1 and SRII2 models. Our PCA analysis found that $\tau=0.0590 \pm 0.0024$ for the V1 model and $\tau=0.0591 \pm 0.0025$ for the SRII2 model. The input values are $\tau=0.05998$ and $\tau=0.05960$ for the V1 and SRII2 models, respectively, and thus we successfully recovered the input $\tau$ values. Note that in the PCA analysis $\tau$ and $\tft$ are not fundamental parameters, but derived parameters. 

We should caution, however, that the reionization history reconstructed by Eq.~(\ref{eq:eq1}) using the estimated PCs from the MCMC analysis can be different from the input reionization history. The PCA method only guarantees that the $C_\ell$s calculated using recovered PCs should be indistinguishable from the input $C_\ell$s within the statistical uncertainty assumed in the analysis.

In the PCA analysis where $m_\mu$ are fundamental parameters, some combinations of $m_\mu$ break the physicality condition and $x_e$ becomes negative, and hence $\tau$ and $\tft$ can take the negative values. Therefore we have finite probabilities in the negative $\tft$ region.

We also perform the MCMC analysis with the PCA parameters in the logarithmic basis introduced in Eqs~(5) and (6). In Figure~\ref{fig:contours_log_PCA}, we show the triangle plots for the posterior distributions for the PCA parameters. We found that the results are strongly affected by the prior volume effect; and in fact, some of the input models are outside the 2$\sigma$ contours especially for $m_1$ parameter. 

As we mentioned earlier, we also varied the standard six cosmological parameters. We found that these parameters were tightly constrained by the temperature power spectrum of the CMB and their input values can be recovered unbiased irrespective to the reionization models considered here.

\subsection{FlexKnot analysis}
\label{sub:FlexKnotresult}
FlexKnot and FlexKnot-m show similar behaviors in estimating $\tau$. Both for the mock data corresponding to SRII1 and V1, these methods probe $\tau$ in a biased way such that the estimated $\tau$ is larger than the true $\tau$ (Figure \ref{tau_tougou}). Compared to the PCA analysis, the peak value of $\tau$ in the marginalized posterior distribution is shifted by $\sim 0.003$. Compared to the true value, the peak value of $\tau$ is shifted by $\sim 0.0025$. The deviation of the posterior is $\sim 10 \%$ smaller than that of the PCA. 

Such a bias of FlexKnot and FlexKnot-m in estimating $\tau$, against PCA, has also been reported in MB and \citet{PlanckCollaboration2018}. For PLD, the estimated $\tau$ from FlexKnot is larger than that from PCA, regardless of the prior chosen (flat-$\tau$ prior or flat-knot prior) and the range of samples ($TT$+$TE$+$EE$ on all $\ell$ or low-E-mode only)  \citep{PlanckCollaboration2018}. Note that the prior we use in this work for both FlexKnot and FlexKnot-m is the flat-knot prior. We also note that the tendency of the flat-$\tau$ prior FlexKnot is to estimate $\tau$ somewhere in between PCA and the flta-knot prior FlexKnot \citep{PlanckCollaboration2018}. If we assume the same tendency, there is the possibility that FlexKnot and FlexKnot-m will better estimate $\tau$ for a CV-limited CMB apparatus.

We find that CV-limited experiments have potential to probe the high-redshift ($z\gtrsim 15$) reionization history. Estimation of $\tft$ shows a similar trend in the peak value, while a significantly different trend in the dispersion (Figure \ref{tau15_tougou}). The peak values of the estimated $\tft$ are slightly smaller in FlexKnot and FlexKnot-m than in PCA, but they are all mutually consistent at $\lesssim 1\sigma$ level of FlexKnot and FlexKnot-m. Very importantly, FlexKnot and FlexKnot-m can provide a relatively tight, two-tailed posterior for $\tft$. SRII2, a reionization history with high-redshift ionization tail with $\tft=0.008$ can be distinguished from V1, a reionization history with practically null ionization tail with $\tft=0.0004$. FlexKnot-m stands out in providing a tighter constraint on $\tft$ than FlexKnot, and the distinction of SRII2-likes from V1-likes is the strongest (at $\gtrsim 2\sigma$ level as long as $\tft \gtrsim \tft({\rm SRII2})-\tft({\rm V1})=0.0075$). Therefore, contrary to the claim by \citet{Wu2021}, CV-limited experiments can tell whether high-redshift ionization tail exists or not, with a decent significance.

The real problem is that, even though CV-limited experiments can probe a high-redshift ionization tail if substantial, the estimated $\tft$'s by FlexKnot and FlexKnot-m are in general smaller than the true value if $\tft\sim 0.008$, and larger than the true value if $\tft\lesssim 4\times 10^{-4}$ (Figure \ref{tau15_tougou}). This indicates that either the CMB E-mode is only weakly sensitive to $\tft$, or FlexKnot and FlexKnot-m are not the most optimal methods to probe $\tft$. In comparison, PCA estimates the peak $\tft$ at a value closer to the true value for SRII2 mock and at a similar value with FlexKnot for V1 mock. However, PCA yields a much larger deviation in the marginalized posterior and encompasses non-physical quantities, $\tft<0$. Therefore, FlexKnot and FlexKnot-m seem superior to PCA in estimating $\tft$.

\subsection{HS17 analysis}
\label{HS17result}
We find that HS17 analysis is overall similar to FlexKnot and FlexKnot-m in the estimative power for $\tau$. As shown in Figure \ref{tau_tougou}, HS17 estimates $\tau$ in a way biased toward larger values than that by PCA. HS17-3 shows almost identical behavior with FlexKnot in estimating both $\tau$ and $\tft$ regardless of the existence of the high-redshift ionization tail. HS17-7, on the other hand, estimates both $\tau$ and $\tft$ in the most biased way: estimated values of these derived parameters are the largest amongst all the methods.
This fact indicates that if the true reionization history were the types described by usual reionization models, where $\xe$ grows monotonically in time at $z>15$, having as many as seven knots in the analysis would be redundant and could cause the unwanted bias.

\begin{figure*}[htb!]
    \centering
    \includegraphics[width=0.32\textwidth]{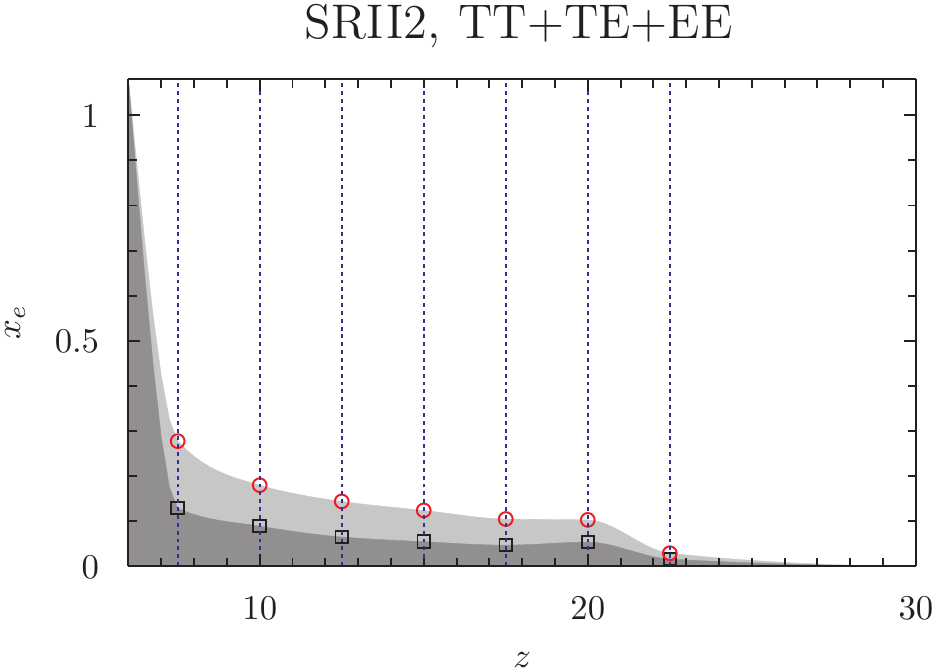}
    \includegraphics[width=0.32\textwidth]{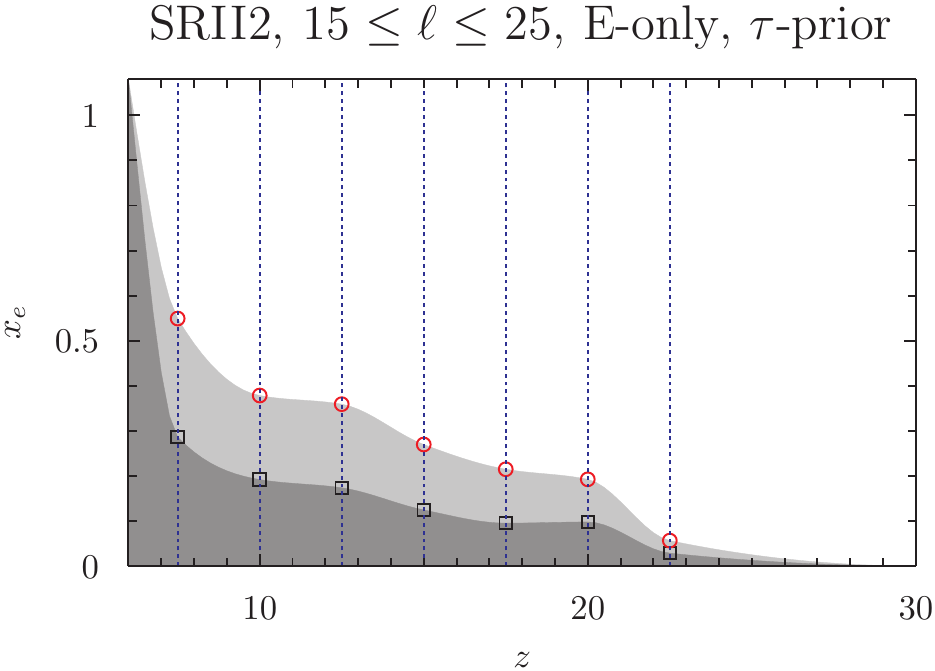}
    \includegraphics[width=0.292\textwidth]{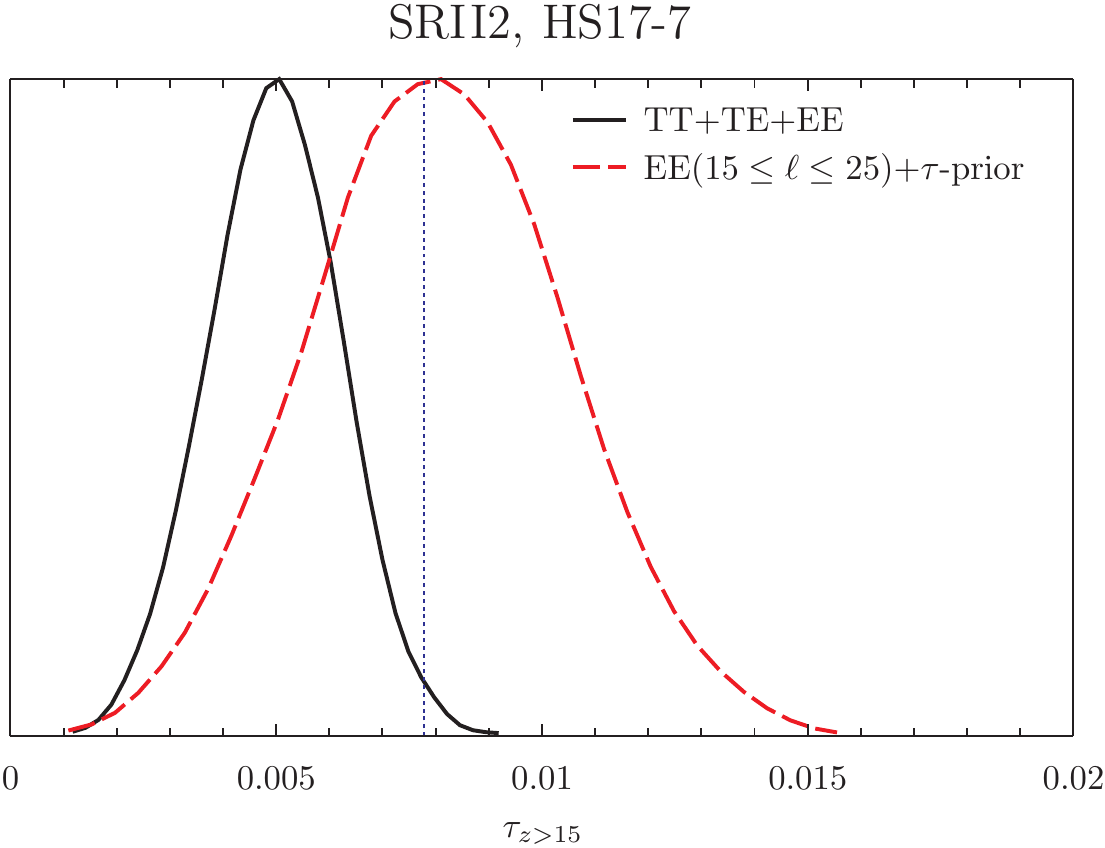}
    \caption{Illustration on how resampling the CMB data can change the estimate on $\xe (z)$ and $\tft$. ({\em left}) Constraints on $\xe(z)$ by the HS17-7 analysis on the full $TT$+$TE$+$EE$ mock data based on SRII2.  68 \% (dark grey) and 95 \% (light grey) constraints on $\xe(z)$  are constructed by connecting the individual marginalized posteriors of $\xe(z_{i})$ at 68 \% (square) and 95 \% (circle) confidence levels, respectively, with the PCHIP interpolation. ({\em middle}) Constraints on $\xe(z)$ by the HS17-7 analysis on the $EE$-only mock data based on SRII2, limited to $15\le\ell\le 25$, and with the $\tau$-prior. ({\em right}) Marginalized posteriors of $\tft$, corresponding to the full $TT$+$TE$+$EE$ analysis (black solid) and the resampling analysis with only $EE$, $15\le\ell\le 25$, and the $\tau$-prior. The vertical line denotes the true value of $\tft$ in SRII2.}
    \label{fig:HS17_zxe_tau15}
\end{figure*} 

\begin{figure*}[htb!]
    \centering
    \includegraphics[width=0.32\textwidth]{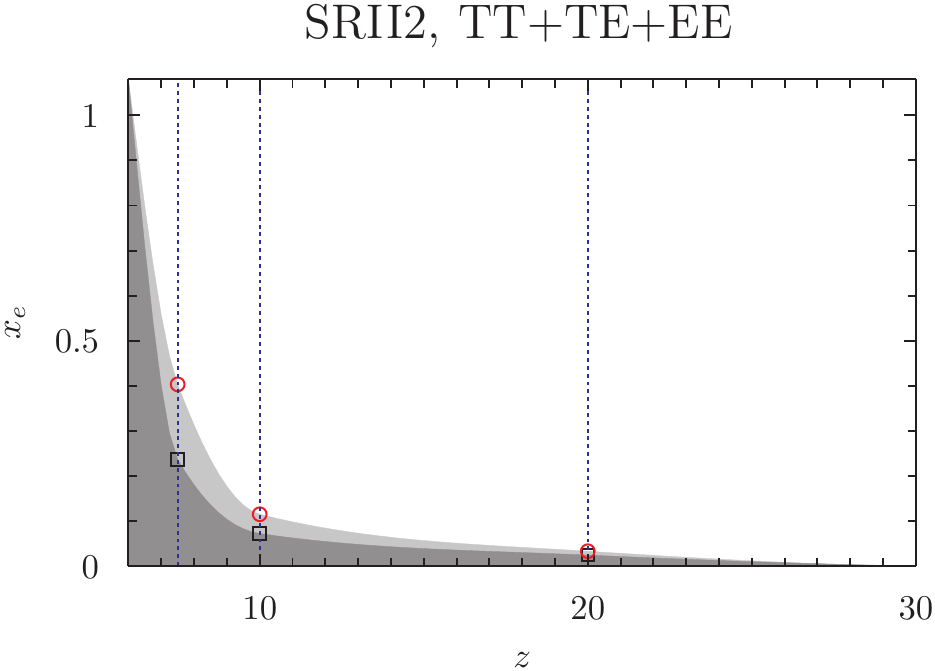}
    \includegraphics[width=0.32\textwidth]{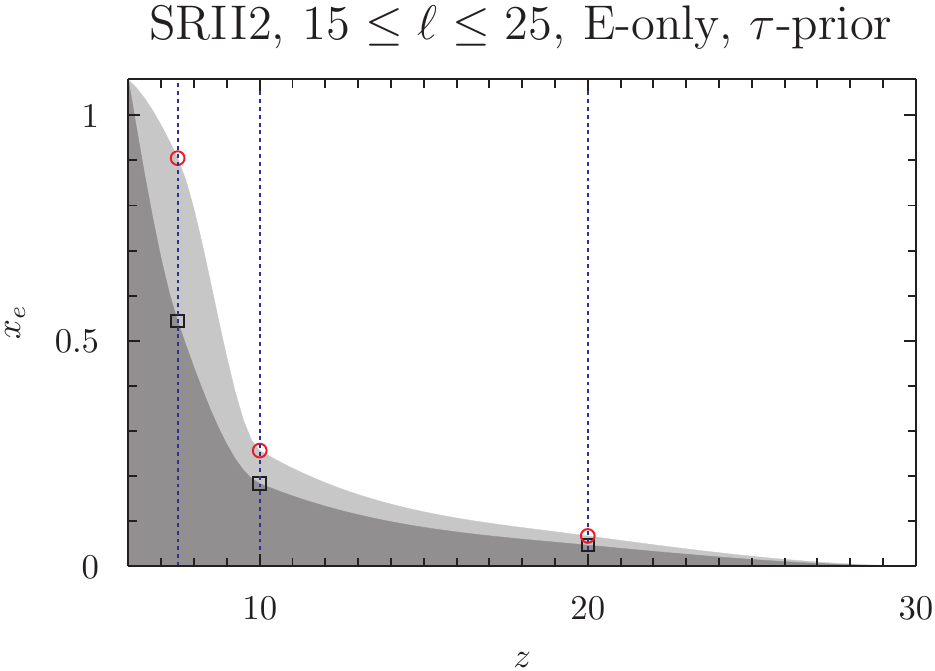}
    \includegraphics[width=0.292\textwidth]{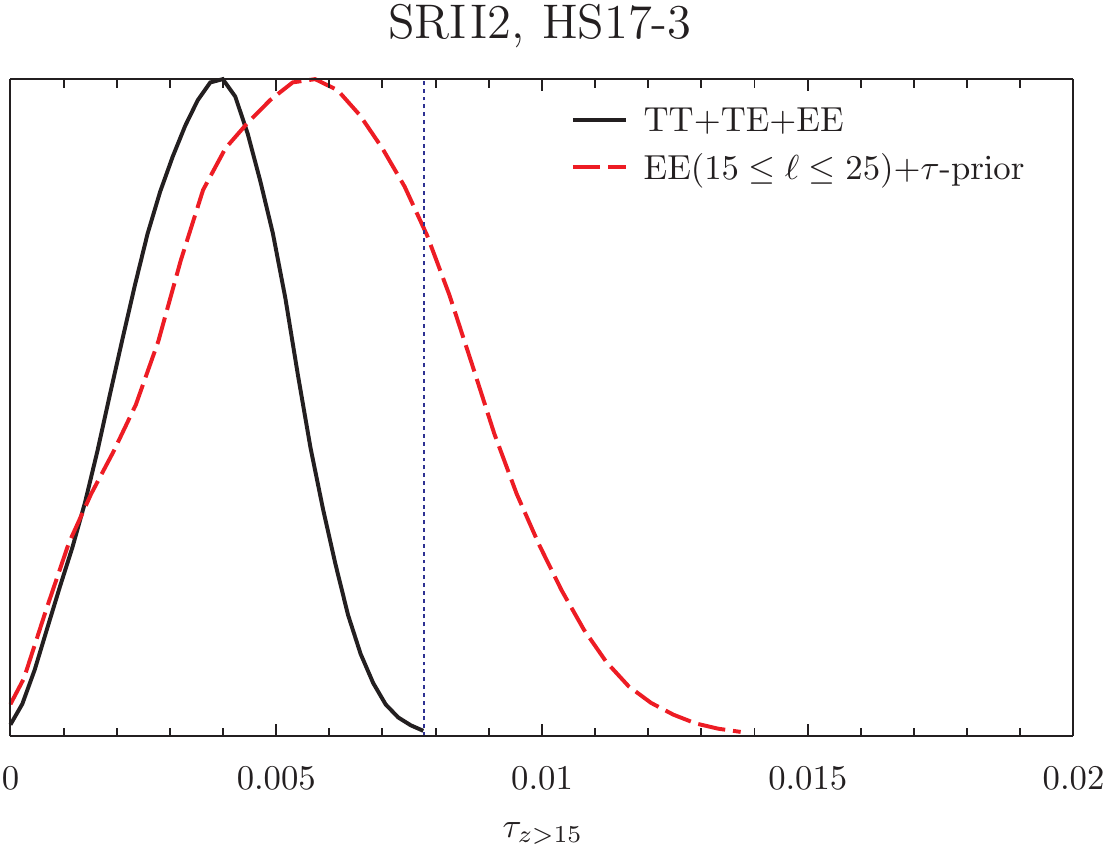}
    \caption{Same as Figure \ref{fig:HS17_zxe_tau15}, but with HS17-3 analysis.}
    \label{fig:HS17_3_zxe_tau15}
\end{figure*} 

HS17 is also similar to FlexKnot and FlexKnot-m in estimating $\tft$. Both HS17-3 and HS17-7 have the potential to distinguish the high-redshift-tailed model SRII2 from the tail-less model V1. Both HS17-3 and HS17-7 suffer from the bias, similar to FlexKnot and FlexKnot-m, in estimating $\tft$. For SRII2 mock, the HS17-3 estimate of $\tft$ is almost identical to that of FlexKnot, and the HS17-7 estimate of $\tft$ to that of FlexKnot-m. For V1 mock, on the other hand, the HS17-3 estimate of $\tft$ is almost identical to that of FlexKnot-m, and then HS-17 gives the largest estimate of $\tft$ (Figure \ref{tau15_tougou}). Nevertheless, these differences are very small for the SRII2 mock, and mutually consistent within $\lesssim 1\sigma$ for the V1 mock. The remaining question is whether one could benefit from the resampling method (Section \ref{subsec:tau15}) in estimating $\tft$, which will be described in Section \ref{subsec:result_tau15}.

\subsection{resampling CMB E-mode with prior on the optical depth and limited spherical harmonics}
\label{subsec:result_tau15}
Now we describe the result of our resampling analysis (Section \ref{subsec:tau15}) with the likelihood given by equation (\ref{chi1525Eonlytauprior}). The intention of such a resampling is to obtain a better estimate of $\tft$ than the full analysis with equation (\ref{eq:chi2}), because the result of the latter yields biased estimates of $\tft$ as described in Sections \ref{subsec:PCA} -- \ref{subsec:HS17}. The resampling definitely changes the estimated $\xe (z)$ and $\tft$, as illustrated in Figures \ref{fig:HS17_zxe_tau15} and \ref{fig:HS17_3_zxe_tau15} for the cases of HS17-7 and HS17-3 when the true model is SRII2.

The result is best summarized in Figures \ref{tau15_tougou_E} and \ref{tau15_collection}, and is conflicted in terms of reaching the original intention. Against SRII2 mock, all sampling methods with equation (\ref{chi1525Eonlytauprior}), except PCA, yield the peak $\tft$ values boosted (red points in Figure \ref{tau15_collection}) from those (black points in Figure \ref{tau15_collection}) with equation (\ref{eq:chi2}). FlexKnot-m and HS17-7 stand out in terms of yielding the peak $\tft$ that is almost identical to the true value (dotted line). It is then tempting to choose these two methods over others to probe $\tft$ most accurately when the Universe were represented by SRII2 types. However, this tendency to boost estimated $\tft$ remains even for V1 that has practically null $\tft$ (but also with the exception of PCA). As a result, FlexKnot-m and HS17-7, which stood out in estimating the $\tft$ of SRII2 accurately, now estimate $\tft$ of V1 mock to be at $\sim 0.04$--$0.05$. The change of peak $\tft$'s in FlexKnot and HS17-3 are relatively milder than FlexKnot-m and HS17-7. Especially, HS17-3 has a substantial boost of the peak $\tft$ for SRII2, larger than $1\sigma$, but a minor boost for V1. Therefore, HS17-3 works most favorably if one is to single out a method that works in most general cases. Still, the difference of the peak $\tft$ of HS17-3 from the true value is somewhat disappointing.

The overall increase of deviation in $\tft$ for the resampling also goes against our intention (Figure \ref{tau15_collection}). This seems to imply that the impact of $\tft$ on $\ClEE$ is not completely limited to $15\le\ell\le 25$, and thus one may have to include lower $\ell$ data to some extent, maybe with some weighting, to reduce the deviation in $\tft$. The cause of the increase of this $\tft$-deviation in the resampling does not seem to caused by the exclusion of the $TT$ and $TE$ power spectrums: we experimented using all of the $TT$, $TE$ and $EE$ modes with limited-$\ell$, and the resulting deviation in $\tft$ is almost the same as in the $EE$-only, limited-$\ell$ case.

Therefore, we conclude that one cannot choose a single method that achieves the original intention of correctly estimating $\tft$. Still, it is tempting to use FlexKnot-m or HS17-7 with the $\tau$-modulated, limited-$\ell$ likelihood (equation \ref{chi1525Eonlytauprior}) when the peak $\tft$ is estimated to be $\gtrsim 0.004$ with the full $TT$+$TE$+$EE$ analysis. Also, one might want to take the face value of $\tft$ instead and do not further perform this resampling, when $\tft$ is estimated to be $\lesssim 0.002$ with the full $TT$+$TE$+$EE$ analysis. Nevertheless this is not highly recommended, and we need to find a more self-consistent method to probe $\tft$.  

Amongst all, PCA works the most poorly in estimating $\tft$ both in its peak value and deviation of the posterior. For SRII2 mock the peak $\tft$ is now close to the null, and for V1 mock the peak $\tft$ is negative. The deviation is about 2--3 times as large as that in the full $TT$+$TE$+$EE$ analysis. Therefore, equation (\ref{chi1525Eonlytauprior}) does not work as intended for PCA, and unless physicality condition is enforced, PCA would not benefit from using equation (\ref{chi1525Eonlytauprior}).

\section{Summary and Discussion}
\label{section:summary}
We investigated whether the CMB angular power spectrums of the temperature and the polarization, measured by an ideal CV-limited apparatus such as LiteBIRD, would provide a hint on the early stage of reionization. Toward this end, we first generated CMB mock data based on a set of reionization models in which the contribution by a wide range of halo species is carefully treated. Then, we selected various methods that sample the reionization histories during the MCMC: PCA, FlexKnot and HS17. Also, we specified the CMB likelihood not only in the standard way (using the full-$\ell$ range in the $TT$, $TE$ and $EE$ power spectrum) but also in a "resampled" way (using a limited-$\ell$ range in the $EE$ power spectrum, with the prior condition on $\tau$ enforced).

\begin{figure}
    \centering
    \includegraphics[width=8cm]{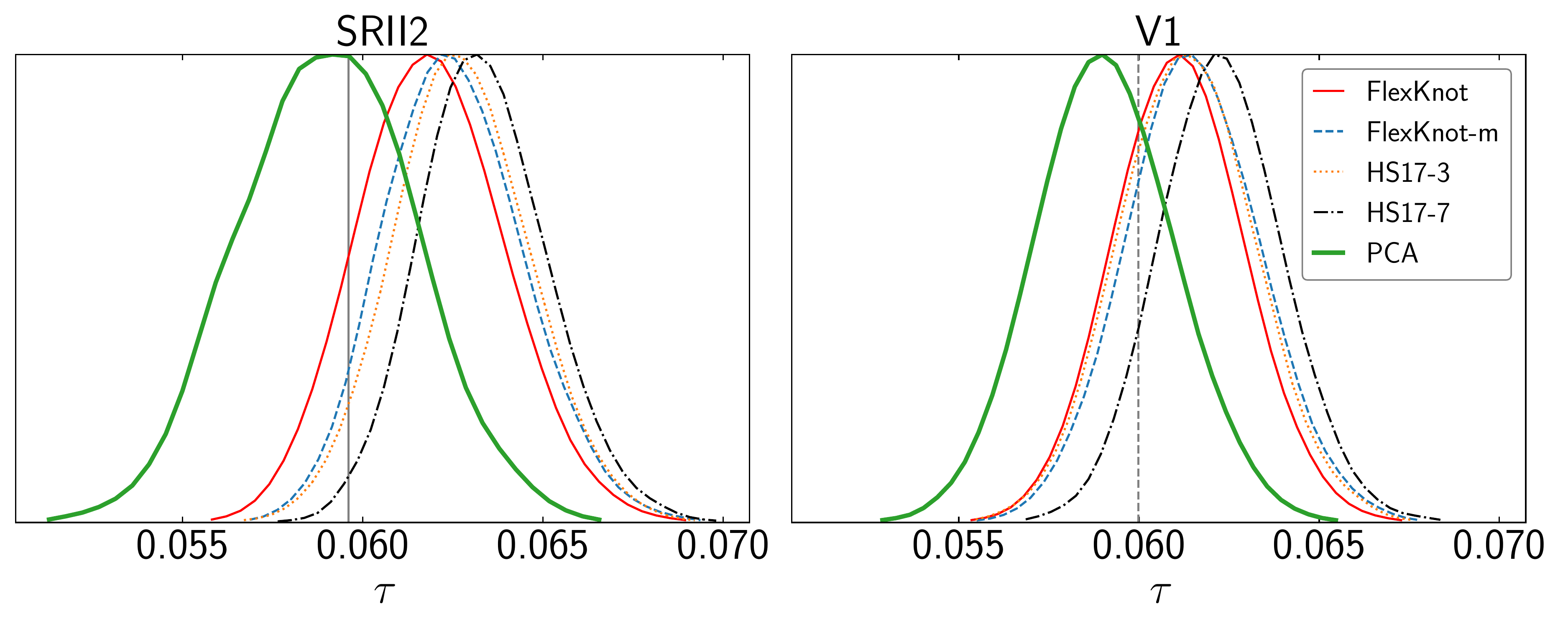}
    \caption{left panel: Marginalized distributions of $\tau$ for different sampling methods (FlexKnot: red, thin solid; FlexKnot-m: blue, dashed; HS17-3: orange, dotted; HS17-7: black, dot-dashed; PCA: green, thick sold) given a mock data based on the SRII2 model, obtained from analyzing all the CMB power spectra ($TT$, $TE$, $EE$) when $f_{\rm sky} = 1$. The vertical lines are the true values of $\tau$ (see table \ref{tab:models}). right: Same as the left panel, but with a mock data based on the V1 model.}
    \label{tau_tougou}
\end{figure}

\begin{figure}
    \centering
    \includegraphics[width=8cm]{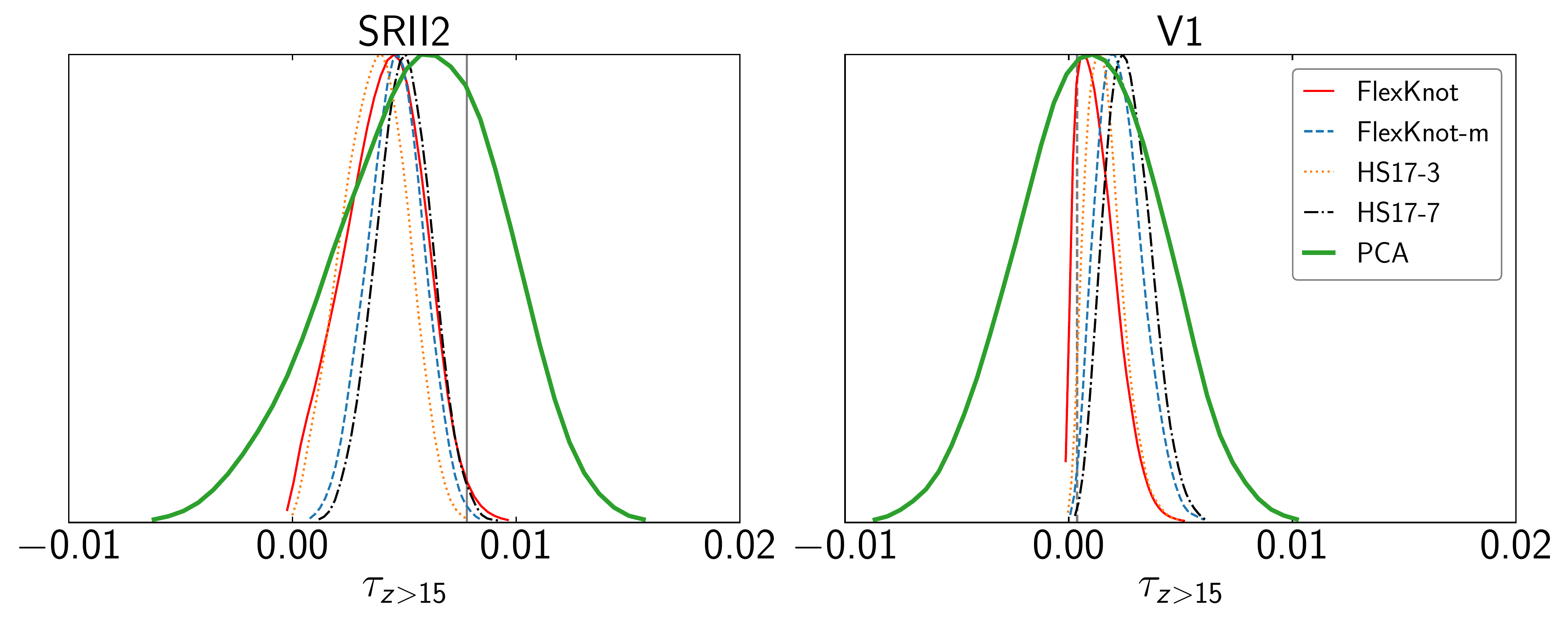}
    \caption{Same as Figure \ref{tau_tougou}, except that marginalized distributions of $\tft$ for different sampling methods}
    \label{tau15_tougou}
\end{figure}

\begin{figure}
    \centering
    \includegraphics[width=8cm]{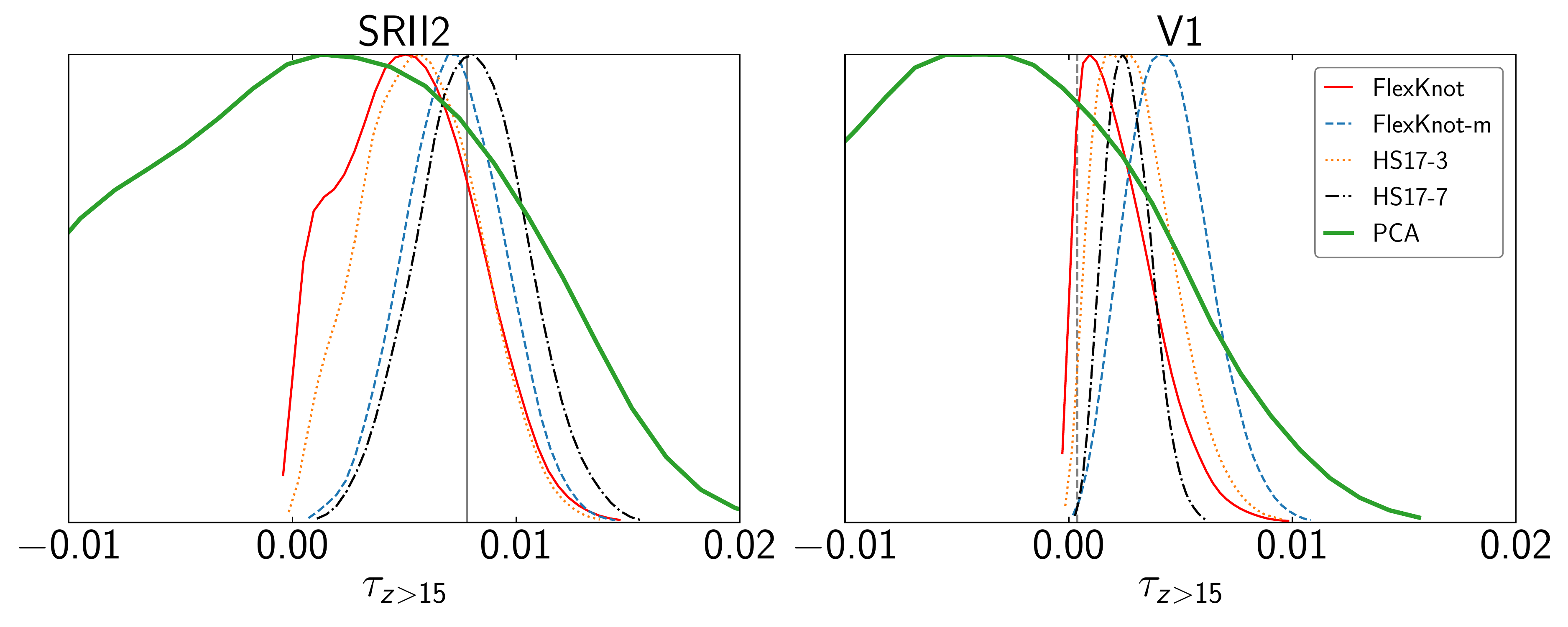}
    \caption{Estimation of $\tft$ by resampling CMB E-mode with the limited range of $\ell$ and the $\tau$ prior, using the likelihood given by equation (\ref{chi1525Eonlytauprior}) for SRII2 mock (left) and V1 mock (right).The vertical lines are the true values of $\tau$. Line conventions follow those of Figures \ref{tau_tougou} and \ref{tau15_tougou}.}
    \label{tau15_tougou_E}
\end{figure}

\begin{figure*}
    \centering
    \includegraphics[width=8cm]{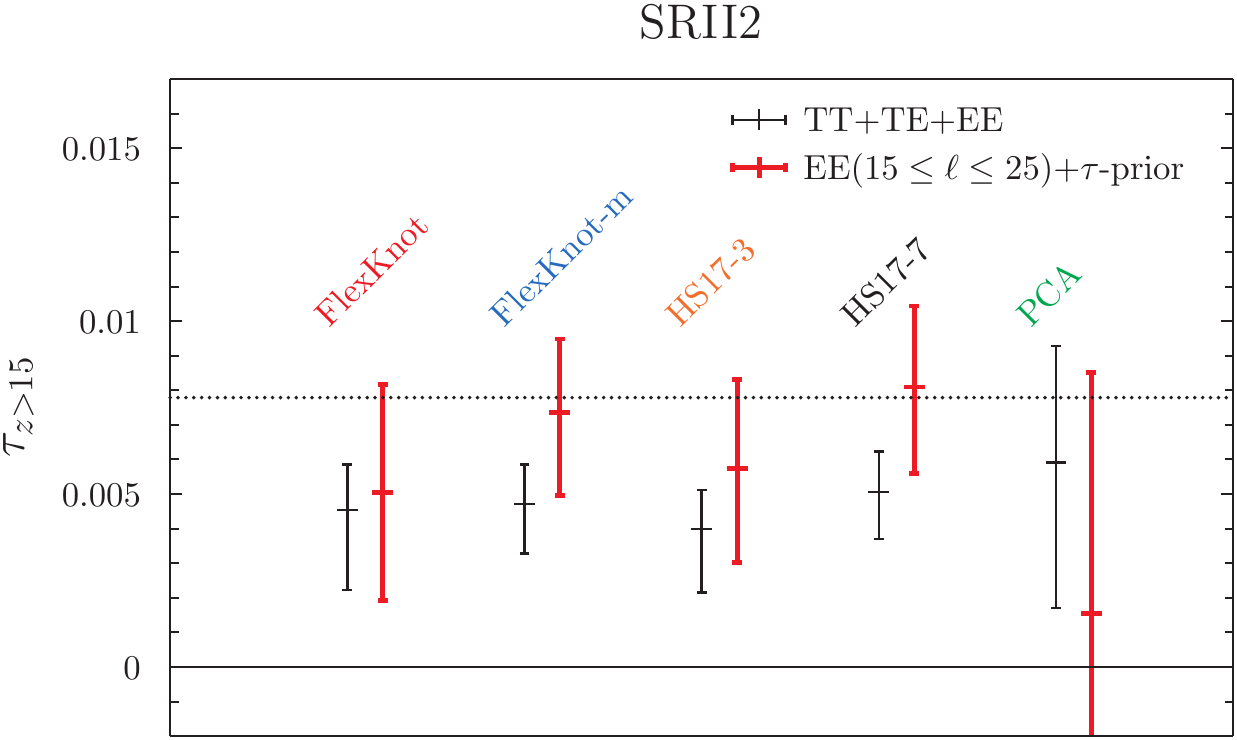}
    \includegraphics[width=8cm]{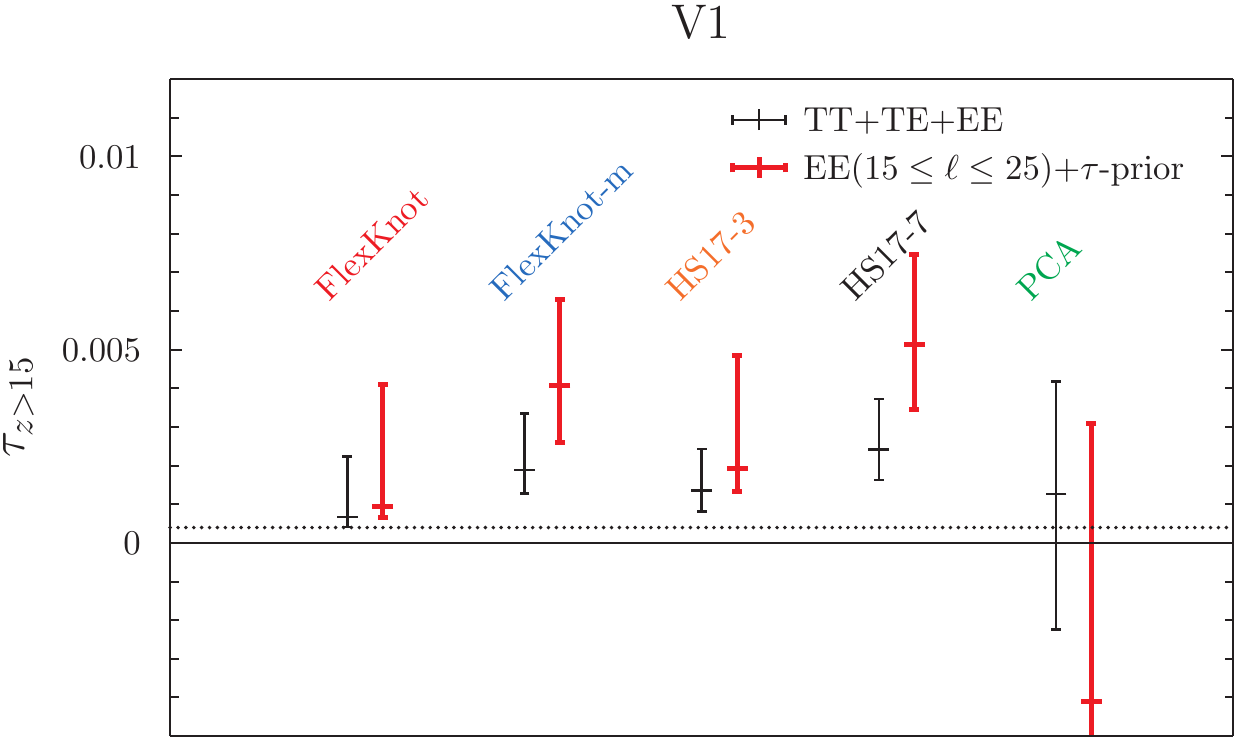}
    \caption{Performance of sampling methods for $\tft$ estimation, seen in the peak values (point within each error bar) with error bars that correspond to 68\% of the posteriors shown in Figure \ref{tau_tougou} (thin, black) and \ref{tau15_tougou} (thick, red). These are plotted against the true values of $\tft$ for SRII2 (left) and V1 (right), represented by dotted lines. Pairs of points represent, from left to right, the FlexKnot, FlexKnot-m, HS17-3, HS17-7, and PCA.}
    \label{tau15_collection}
\end{figure*}

All sampling methods show reasonable estimates on $\tau$, of course with tighter constraints than the Planck estimate by a CV-limited experiment when the usual, standard likelihood is used. A CV-limited experiment is also much more powerful than the Planck in estimating the partial optical depth $\tft$, which is a clear indicator on the early history of reionization. Such an experiment provides two-tailed distribution for the marginalized posterior of $\tft$ when $\tft\sim 0.008$, and is expected to do so even for smaller $\tft$. Therefore, CV-limited experiments will be able to tell whether the high-redshift ($z\gtrsim 15$) reionization was active or not. However, we find that estimates on the partial optical depth $\tft$ are biased toward smaller (larger) values than the true ones when the base model for the mock data has a significant (negligible) $\tft$.

When the resampled likelihood is used, on the other hand, models with active reionization at $z>15$ seem to be better probed such that the peak of the $\tft$ posterior becomes much closer to the true value. While this is promising, with this likelihood, models with negligible reionization at $z>15$ are now more poorly probed. In this work, we could not find a single method that consistently probes $\tft$ accurately. At the point, therefore, we conclude that an accurate estimation of $\tft$ remains somewhat elusive. Nevertheless, even in the framework of a single data (CMB anisotropy), such a resampling bears some prospect in estimating $\tft$ accurately when high-redshift reionization is active. Future work to find out an optimal and consistent way to probe $\tft$ is warranted.

Each sampling method has pros and cons. PCA has a reasonable predictive power on $\tau$ but suffers from non-physicality in $\xe (z)$, and performs most poorly in estimating $\tft$ among the tested sampling methods. FlexKnot, FlexKnot-m, HS17-3 and HS17-7 have more or less similar  behavior in estimating $\tau$ and $\tft$. However, discriminating different reionization histories with these methods are not as straightforward as PCA. The reason may be because principal components are weighted and sorted in its relative importance to $\ClEE$ and thus the major two ($m_1$, $m_2$) or three ($m_1$, $m_2$, $m_3$) components determine the most part of a reionization history. On the contrary, HS17 samples $\xe(z)$ evenly for each $i$th knot and thus the discrimination of different reionization models should be performed on the $z$--$\xe$ plane. However, we find it difficult to use marginalized $\xe(z)$'s for model discrimination because the marginalized posterior distribution of any $\xe(z)$ we find is usually one-sided (see that there are only upper contours in Figures \ref{fig:HS17_zxe_tau15} and \ref{fig:HS17_3_zxe_tau15}). This does not translate to, however, the incapacity to probe $\tft$ but indeed $\tft$ posterior of HS17 is double-sided. This symptom, having one-sided $\xe(z_{i})$ for $i$th knot, exists also in FlexKnot as long as $z_{i}>15$. So at this point, we find it difficult to use HS17 or FlexKnot for direct discrimination of reionization histories, and one may have to use $\tft$ as the only parameter for such a discrimination. We will study to better constrain reionization histories in the future.

As we mentioned in Sec.~\ref{subsec:PCA}, the total optical depth parameter $\tau$ and the partial one $\tft$ are not fundamental parameters but derived ones. We find that the posterior distribution of $\tau$ can recover the input values almost accurately in PCA (Figure \ref{tau_tougou}). In case of HS17 and FlexKnot, as mentioned earlier, distributions have smaller variances than those derived from the PCA, but they all tend to show larger $\tau$ values, where the biases are around $1$-$2\sigma$ levels. We believe that this is the effect of the different prior. For example, the original FlexKnot has the flat prior in $z$-$\xe$ plane, and its bias on $\tau$ may be suppressed if the flat-$\tau$ prior is enforced as \citet{Millea2018} has proposed. Nevertheless, there is no firm motivation to believe the flat prior for a derived parameter such as $\tau$, and this issue should be instead tackled by forward modelling as we did here: one generates a mock data from a theory (set of theories), and observe how any specific MCMC method and prior condition perform in estimating a targeted, derived parameter.

As for PCA, violation of the physicality is always an issue. We find that the logarithmic-basis PCA takes care of the non-physicality but instead the predictive power we have seen from the linear-basis PCA is lost. This can be considered as another example of the effect of the prior condition. Going back to the linear-basis PCA, then, the predictive power on $\tft$ is much poorer than HS17 and FlexKnot. This is partly because of non-physicality, where non-physicality in $\xe$ is relatively more severe at $z>15$ than at $z<6$. One could enforce physicality during MCMC sampling as \citet{Ahn2012}. 

In this study we assumed a cosmic variance limited observation of the $\ClEE$ power spectrum, but this would not be an ultimate constraint on $\tau$ from the CMB.  \citet{2018PhRvD..97j3505M} argued that one may obtain even more precise constraint on $\tau$ using polarization signal induced by free electrons in galaxy clusters. Since such polarization signal is determined by the remote CMB quadrupoles viewed by the galaxy clusters, and since these quadrupoles induce the CMB E-mode polarization on large scales, cross-correlating these maps can be used to improve the constraint on $\tau$ by an order of magnitude over the constraint from the cosmic variance limited E-mode power spectrum alone. This polarization signal has not yet been detected, but given its statistical power as a cosmology probe, it is worth a further investigation.

Among many reionization-related observations, the claimed detection of the absorption against the continuum foreground by the Experiment to Detect the Global EoR Signature (EDGES; \citealt{EDGES}) is the most relevant one to the early phase of the cosmic reionization. In terms of the 21-cm-line brightness temperature, $T_{b}\simeq -500\,{\rm mK}$, whose absolute value is unreachable from the absolute maximum of $\sim 200\,{\rm mK}$ in the standard $\Lambda$CDM cosmology. Such a large value of absorption brings in the possibility of exotic scenarios such as the existence of a foreground of yet unknown origin \citep{EwallWice_foreground} or the unconventionally high interaction of dark matter particles with baryons \citep{Tashiro_DMbaryon,Barkana_DMbaryon}. Now, the new observation by the Shaped Antenna measurement of the background RAdio Spectrum 3 (SARAS 3) seems to contradict the EDGES analysis, with null detection of the claimed signal of EDGES \citep{SARAS3}. Having an independent constraint on the physics at $z>15$ through the future CMB observation, therefore, would become useful in determining the fate of the $\Lambda$CDM paradigm, judging the interpretations of the EDGES and the SARAS 3 data, and understanding how early astrophysical radiation sources emerged.

\section*{Acknowledgements}
This work is supported in part by the JSPS grant numbers 18K03616, 17H01110, 21H04467 and JST AIP Acceleration Research Grant JP20317829 and JST FOREST Program JPMJFR20352935. KA is supported by NRF-2016R1D1A1B04935414, 2021R1A2C1095136, and 2016R1A5A1013277. KA also appreciates APCTP and KASI for their hospitality during completion of this work. This study was supported by research fund from Chosun University, 2020.
We thank A. Chatterjee for carefully reading the draft and giving comments.

\label{lastpage}

\bibliography{paper}{}
\bibliographystyle{aasjournal}

\end{document}